\documentclass[useAMS,usenatbib]{mnras}

\usepackage{color}
\usepackage[english]{babel}
\usepackage{graphicx}
\usepackage{amsmath}
\usepackage{amssymb}
\usepackage{epsf}
\usepackage{bm}
\usepackage[utf8]{inputenc}
\begin{document}

\def\pFn{p_{\rm Fn}}
\def\pFp{p_{\rm Fp}}
\def\pFe{p_{\rm Fe}}

\newcommand{\om}{\mbox{$\omega$}}              
\newcommand{\Om}{\mbox{$\Omega$}}              
\newcommand{\Th}{\mbox{$\Theta$}}              
\newcommand{\ph}{\mbox{$\varphi$}}             
\newcommand{\del}{\mbox{$\delta$}}             
\newcommand{\Del}{\mbox{$\Delta$}}             
\newcommand{\lam}{\mbox{$\lambda$}}            
\newcommand{\Lam}{\mbox{$\Lambda$}}            
\newcommand{\ep}{\mbox{$\varepsilon$}}         
\newcommand{\ka}{\mbox{$\kappa$}}              
\newcommand{\dd}{\mbox{d}}                     
\newcommand{\nb}{\mbox{$n_{\rm b}$}}                     

\newcommand{\Tb}{\mbox{$T_{\rm b}$}}         
\newcommand{\kB}{\mbox{$k_{\rm B}$}}           
\newcommand{\tn}{\mbox{$T_{{\rm c}n}$}}        
\newcommand{\tp}{\mbox{$T_{{\rm c}p}$}}        
\newcommand{\te}{\mbox{$T_{eff}$}}             
\newcommand{\ex}{\mbox{\rm e}}                 
\newcommand{\rate}{\mbox{${\rm erg~cm^{-3}~s^{-1}}$}}
\newcommand{\gcc}{\mbox{${\rm g~cm^{-3}}$}}
\newcommand{\xr}{\mbox{$x_{\rm r}$}}
\newcommand{\gs}{\mbox{$g_{\rm s}$}}
\newcommand{\me}{\mbox{$m_{\rm e}$}}
\newcommand{\xg}{\mbox{$x_{\rm g}$}}
\newcommand{\msun}{\mbox{${\rm M}\odot$}}
\def\dy{\textcolor{red}}
\def\mg{\textcolor{magenta}}
\def\ak{\textcolor{green}}



\title[Selfsimilar torsional oscillations of neutron stars]{Selfsimilarity relations for 
torsional oscillations of neutron stars}

\author[D. G. Yakovlev ]{{D. G. Yakovlev\thanks{E-mail: yak.astro@mail.ioffe.ru}},\\
    Ioffe Institute, Politekhnicheskaya 26, St~Petersburg 194021, Russia\\
    }

\maketitle \label{firstpage}

\begin{abstract} 
Selfsimilarity relations for torsional oscillation frequencies
of neutron star crust are discussed. For any neutron star model, 
the frequencies of fundamental
torsional oscillations (with no nodes of radial 
wave function, i.e. at $n=0$, and at all possible 
angular wave numbers $\ell \geq 2$) is determined by a single constant.
Frequencies of ordinary torsional oscillations (at any
$n>0$ with $\ell \geq 2$) are determined by two constants.
These constants are easily calculated through radial integrals
over the neutron star crust, giving the simplest method to determine full oscillation spectrum. All constants for a star of fixed mass
can be accurately interpolated for stars of various masses (but 
the same equation of state). 
In addition, the torsional oscillations can be accurately studied in 
the flat space-time approximation within the
crust. The results can be useful for investigating
magneto-elastic oscillations of magnetars 
which are
thought to be observed as quasi-periodic oscillations after flares of
soft-gamma repeaters.
\end{abstract}



\section{Introduction}
\label{s:introduc}

Neutron stars consist of massive cores and thin light envelopes
\citep[e.g.,][]{ST1983}. The core is liquid and contains
superdense matter which equation of state (EOS) is still
not entirely known. The envelope is only $\sim 1$ km thick and
has the mass $\sim 0.01\,\msun$. It consists of 
electrons, atomic nuclei, and (at densities $\rho$
higher the neutron drip density $\rho_{\rm drip} \approx 4.3 \times 10^{11}$ \gcc)  quasi-free neutrons. As a rule, the atomic nuclei constitute 
Coulomb crystal (e.g. \citealt{HPY2007}) which 
melts only in the very surface layers of the star. 
 {The} envelope is often called the crust that is divided into
the outer ($\rho< \rho_{\rm drip}$) and the inner
($\rho > \rho_{\rm drip}$) crust. {The maximum density in the
crust ( {$\rho_{\rm cc}\sim 1.4 \times 10^{14}$ \gcc}) is limited by the core.}

This paper is devoted to the theory of torsional oscillations of
neutron star, in which case only the crystallized shell
oscillates, while other layers do not. The star is assumed to be 
non-magnetic and spherical. Foundation of the theory
was laid by \citet{1980Hansen,1983ST} and \citet{1988McDermott}.
Later developments were numerous (see, e.g., \citealt{2007Samuel,2009Andersson,2012Sotani,2013aSotani,2013Sotani,2016Sotani,2017Sotani,2017aSotani,2018Sotani,2019Sotani}
and references therein).

The theory started new life after the discovery 
of quasi-periodic oscillations (QPOs) in spectra of soft-gamma repeaters (SGRs) after giant flares (see \citealt{2005Israel,2006Watts,2011Hambaryan,2014Huppen,
2014Huppenkothen,2018Pumpe}).
SRGs are magnetars possessing superstrong magnetic fields
$B \sim 10^{15}$ G (e.g. \citealt{OlKas2014,2015Mereghetti,2017KasB}).
Note that seismic activity of SRGs was predicted by
\citet{1998Duncan}.

Seismology of magnetars has been developed
in numerous publications (e.g., \citealt{2006Levin,2007Levin,2006Glampedakis,2007Sotani,2009CD,2009Colaiuda,2011Colaiuda,2012Colaiuda,2011vanHoven,2012vanHoven,2011Gabler,
2012Gabler,2013Gabler,2013Gabler1,2016Gabler,2018Gabler,
2014Passamon,2016Link,2016Gabler,2018Gabler}).  
The observed QPO frequencies (after flares of SGR 1806--20, 
SGR 1900+14 and SGR J1550--5418) fall in the same range
(from $\sim 10$ Hz to a few kHz), as the expected frequencies 
of torsional oscillations of non-magnetic
stars. This means that asteroseismology of magnetars can be related
to torsional oscillations of non-magnetic stars. In any case the
theory of torsional oscillations is basic for testing more
complicated seismology of magnetars. Many publications (e.g. \citealt{2007Sotani})
devoted to the magnetar seismology were mostly focused on 
standard torsional oscillations.

Nevertheless, the magnetar seiesmology has turned out
to be much richer than the theory of non-magnetic stars. It includes
magneto-elastic oscillations which can be treated as
torsional oscillations in strong magnetic field, but they become
coupled to the neutron star core and possibly to the magnetosphere.
The magnetic field opens also other types of
magnetar oscillations based on propagation of Alfv\'en waves
in the entire star. 

Here we turn to pure torsional oscillations and formulate
selfsimilarity relations which 
simplify calculation and analysys of torsional oscillation
frequencies. Section \ref{s:torsion} outlines the standard
theory. Selfsimilarity relations are discussed in Section
\ref{s:freq}. In Section \ref{s:bsk} they are applied to studying
torsional oscillation spectrum of neutron stars composed
of matter with the BSk21 EOS (as an example). In Section \ref{s:previous} the selfsimilarity relations are applied for
analysing torsional oscillation spectra of neutron stars with
other EOSs using results of previous work. Section
\ref{s:bursts+QPOs} gives a general outlook on 
interpretation of QPOs, and Section \ref{s:conclude}
presents conclusions.

\section{Torsional oscillations}
\label{s:torsion}

\subsection{ {Exact formulation of the problem}}
\label{s:exact}

Theoretical background for studying torsional oscillations of the crust of a non-rotating and 
non-magnetic neutron star is well known
(e.g.  \citealt{1983ST,2007Sotani}). The crustal matter is
treated as a poly-crystal (isotropic solid) 
of the Coulomb lattice of atomic 
nuclei. The vibrating layer extends from some solidification density
just under the stellar surface to the bottom of the crystalline
matter at the interface between the crust and liquid stellar core 
(e.g., \citealt{HPY2007}). Here we study linear
oscillations in curved space-time neglecting  {metric} perturbations (i.e., in the relativistic Cowling approximation).

The mertic in a spherically 
symmetric star can be taken in the standard form
\begin{equation}
 {\rm d s}^2=- {\rm e }^{2\Phi}\,{\rm d}t^2 + {\rm e}^{2\Lambda}\,{\rm d}r^2
  + r^2 ({\rm d}\theta^2 + \sin^2 \theta \, {\rm d}\phi^2),
\label{e:metric}
\end{equation}
where $t$ is Schwarzschild time (for a distant observer), $r$ is
a radial coordinate (circumferential radius), $\theta$ and $\phi$ 
are ordinary spherical angles; $\Lambda$ and $\Phi$ are two metric
functions of $r$. At any $r$ one has
\begin{equation}
   \exp \Lambda (r)= \frac{1}{\sqrt{1-2Gm(r)/(rc^2)}},
\label{e:Lambda}   
\end{equation}
where $m(r)$ is the gravitational mass enclosed within a sphere of radius $r$, 
$G$ is the gravitational constant and $c$ is the speed of light.

Let $R$ be the stellar radius, and $M=m(R)$ be  {the} gravitational stellar mass. Outside the star ($r>R$) one has
\begin{equation}
\exp  {(\Phi(r))}= \exp(- {\Lambda(r)})= \sqrt{1 -r_{\rm g}/r},
\label{e:outsideNS}
\end{equation}
where $r_{\rm g}=2GM/c^2$ is the Schwarzschild radius.

Torsional oscillations  are associated with shear 
deformations of crystallized  matter along 
spherical surfaces  {(to avoid energy consuming radial motions
of matter elements in strong gravity)}. In the linear regime,
 {these oscillations} are not accompanied by perturbations
of density and pressure. Oscillation 
eigenmodes can be
specified by multipolarity $\ell=2,3,\ldots$, 
azimuthal number  {$m_\ell$} ($-\ell \leq m_\ell \leq \ell$), as well as
by the number of radial nodes $n=0,1,\ldots$  
The eigenfrequencies
$\omega=\omega_{\ell n}$ (defined here for a distant observer) are naturally 
degenerate in $m_\ell$  {(because the basic stellar configuration is spherical)}. 

In order to
find the oscillation spectrum, it is sufficient to set $m_\ell=0$.  {This will be assumed hereafter.} 
Then vibrating matter elements move along circles at fixed
$r$ and $\theta$ (only the angle $\phi$ varies). Proper
displacements $u_{ {\ell n}}^\phi$ of matter elements can be written as
\begin{equation}
      u_{ {\ell n}}^\phi=r\,Y_{ {\ell n}}(r)\,\exp({\rm i}\omega_{ {\ell n}} t)b_\ell(\theta),
      \quad b_\ell(\theta)= \, 
      -\frac{\partial}{\partial \theta} P_\ell(\cos \theta),
 \label{e:displace} 
\end{equation}
where $P_\ell(\cos \theta)$ is a Legendre polynomial. 

The dimensionless function 
$Y_{ {\ell n}}(r)$ is the radial part of oscillation amplitude; it 
is real for our problem. A complex oscillating exponent
$\exp({\rm i}\omega_{ {\ell n}} t)$ has standard meaning 
(as a real part);
$b_\ell(\theta)$ describes the $\theta$-dependence 
of the oscillation amplitude. For instance, 
$b_2(\theta)=3\cos \theta\,\sin \theta$. 
At any $\ell$,  vibrational motion vanishes along at
the `vibrational' axis $z$ [because 
$b_\ell(\theta)\propto \sin \theta$].

The equation for $Y_{ {\ell n}}(r)$ reads
\begin{eqnarray}
	&&Y''_{ {\ell n}}  +  \left( \frac{4}{r}+\Phi'-\Lambda'+\frac{\mu'}{\mu} \right) Y'_{ {\ell n}}+ \left[ \frac{\rho+P/c^2}{\mu}\,\omega_{ {\ell n}}^2   {\rm e}^{-2\Phi} \right. 	
	\nonumber \\
	&& \left. 
	-\frac{(\ell+2)(\ell-1)}{r^2}\right]{\rm e}^{2\Lambda}Y_{ {\ell n}}=0.
	\label{e:eqforY}    
\end{eqnarray}
Here $\rho(r)$ and $P(r)$ are,  {respectively,} the density and pressure of
crustal matter, and $\mu(r)$ is the shear modulus. 

In order to determine the pulsation frequencies,  {equatiom (\ref{e:eqforY})} has to be solved
with the boundary conditions $Y'_{ {\ell n}}(r_1)=0$ and $Y'_{ {\ell n}}(r_2)=0$ at both
(inner and outer) boundaries $r_{1}$ and $r_2$ of the crystalline matter. 
Actually, the type of  {the} boundary condition at $r_2$ is unimportant
\citep{2020KY} because torsional oscillations are mostly supported
by the inner crust and dragged from there to the outer crust.

Equation (\ref{e:eqforY}) was obtained by 
\citet{1983ST} in a more general form, including
weak emission 
of gravitational waves. It  {reduces} to (\ref{e:eqforY}) in the
relativistic Cowling approximation that is well justified for
torsional oscillations.

Note that the quantity
\begin{equation}
     v_{\rm s}(r)=\sqrt{\frac{\mu(r)}{\rho+P(r)/c^2}},
\label{e:vsound}     
\end{equation}
in the square brackets of equation (\ref{e:eqforY})
is a local velocity of the radial shear wave as measured by a local observer. The second term in the  {square} brackets is the contribution 
of centrifugal forces.

Solving equation (\ref{e:eqforY}), one finds $Y_{\ell n}(r)$ and
 {desired eigenfrequencies} $\omega_{\ell n}$. The solution is obtained up to
some normalization constant. It is convenient to normalize $Y_{ {\ell n}}(r)$ 
by the outer-boundary value $Y_0$, that determines angular
vibration amplitude of crystallized matter at $r=r_2$. The linear regime assumes $Y_0 \ll 1$.  

 {Equation (\ref{e:eqforY}) is  basic for {\it exact} solution of the formulated linear oscillation problem. It allows one to directly calculate $\omega_{\ell n}$.} 

\subsection{ {Equivalent exact formulation}}
\label{s:exact1}

 {Instead of directly solving equation (\ref{e:eqforY}), let us use} the formal expression for oscillation frequencies
\begin{eqnarray}
	&& \omega_{ {\ell n}}^2= [B_{ {2 \ell n}}+(\ell +2)(\ell-1)B_{ {1 \ell n}}]/A_{ {\ell n}},
	\label{e:omegaBA} \\
	&& A_{ {\ell n}}=\int_{r_1}^{r_2}{\rm d}r\, r^4\, (\rho+P/c^2)\,{\rm e}^{\Lambda-\Phi}|Y_{ {\ell n}}|^2,
	\label{e:omegaA} \\
	&& B_{2  {\ell n}}=\int_{r_1}^{r_2}{\rm d}r\, \mu r^4 {\rm e}^{\Phi-\Lambda}|Y'_{ {\ell n}}|^2,
	\label{e:omegaB2} \\
	&& B_{1  {\ell n}} = \int_{r_1}^{r_2}{\rm d}r\,\mu r^2 {\rm e}^{\Phi+\Lambda}|Y_{ {\ell n}}|^2 .
	\label{e:omegaB1}
\end{eqnarray}
It follows from equation (\ref{e:eqforY}) (e.g.
\citealt{1983ST,2020KY}).   {It is more complicated but it will be helpful for subsequent analysis  {in Section \ref{s:freq}. It is fully 
equivalent to directly solving equation (\ref{e:eqforY})}. This has been checked in
numerical results presented below.}

In addition to vibration frequencies, one can 
study vibrational energy $E_{ {\ell n}}^{\rm vib}$ in a mode
($\ell,\,n$),
\begin{eqnarray}
	&& E_{ {\ell n}}^{\rm vib}= \frac{\mathrm{\pi} \ell (\ell+1)}{2 \ell+1}\,
	\int_{r_1}^{r_2}{\rm d}r\, \left[ \omega_{\ell n}^2 (\rho+P/c^2)\,{\rm e}^{\Lambda-\Phi}
	|Y_{ {\ell n}}|^2 \right.
	\nonumber \\
	&& \left. + \mu r^4 {\rm e}^{\Phi-\Lambda}|Y'_{ {\ell n}}|^2 
	+ \mu (\ell+2)(\ell-1)r^2 {\rm e}^{\Phi+\Lambda}|Y_{ {\ell n}}|^2\right]
	\nonumber \\
	&&= \frac{2\mathrm{\pi} \ell (\ell+1)}{2 \ell+1} A_{ {\ell n}}
	\omega_{\ell n}^2.
	\label{e:energy}
\end{eqnarray}  
The first two lines are taken from \citet{1983ST} and \citet{2020KY};
the third line follows from equation (\ref{e:omegaBA}).

 {According to equations (\ref{e:omegaBA}) and (\ref{e:energy}),
the vibration frequencies and energies can be expressed through
three integral quantities $A_{ {\ell n}}$, $B_{ { \ell n}}$ and $B_{ {\ell n}}$ given by
equations (\ref{e:omegaA})--(\ref{e:omegaB1}).}

Equation (\ref{e:omegaBA}) can be rewritten as
\begin{eqnarray}
&&\omega_{ {\ell n}}=\sqrt{\widetilde \omega_{ {\ell n}}^2+(\ell +2)(\ell-1)
	\delta \widetilde	\omega_{ {\ell n}}^2},
\label{eq:omegasqrt} \\
&&\delta \widetilde \omega_{ {\ell n}}^2=\frac{B_{ {1 \ell n}}}{A_{ {\ell n}}}, \quad \widetilde \omega_{ {\ell n}}^2=\frac{B_{2 {\ell n}}}{A_{ {\ell n}}},
\label{eq:omegasqrt1}	
\end{eqnarray}
where $\widetilde\omega_{ {\ell n}}$ and $\delta \widetilde \omega_{ {\ell n}}$ are two auxiliary frequencies which depend generally
on $\ell$ and $n$. In addition to angular frequencies $\omega_{ {\ell n}}$, one
often needs cyclic frequencies $\nu_{ {\ell n}}=\omega_{ {\ell n}}/(2 \mathrm{\pi})$.
A cyclic frequency for a mode $(\ell,n)$ reads
\begin{equation}
\nu_{\ell n} = \left[  \widetilde{\nu}_{\ell n}^2
+ (\ell+2)(\ell-1) \, \delta \widetilde{\nu}_{\ell n}^2  \right]^{1/2} .     
\label{e:generalnu} 
\end{equation}

\section{Oscillation spectrum and its selfsimilarity}
\label{s:freq}

\subsection{Fundamental and ordinary modes}
\label{s:n=0andn>0}

It is well known that properties of fundamental ($n=0$, no nodes
of $Y_{\ell n}(r)$ at 
$r_1 < r < r_2$) and ordinary ($n>1$, one or more  {radial} nodes) oscillations 
are very different.

In both cases, the oscillations are mostly formed in the inner
crust under the two effects, which are (i) the radial propagation of shear
waves with the velocity (\ref{e:vsound}) of $v_{\rm s}\sim 10^8$ cm s$^{-1}$ and (ii) the meridional propagation with about
the same speed due to centrifugal effect.
In equation (\ref{e:eqforY}) the first and second effects
are described, respectively, by the first and second terms in the square brackets. Although all matter elements oscillate only along 
circles on respective spheres with fixed $r$, vibrational energy 
and momentum are distributed over entire crystalline shell due to shear nature of elastic deformations.

\subsubsection{Fundamental oscillations}
\label{s:n=0}

\begin{figure}
	\includegraphics[width=0.5\textwidth]{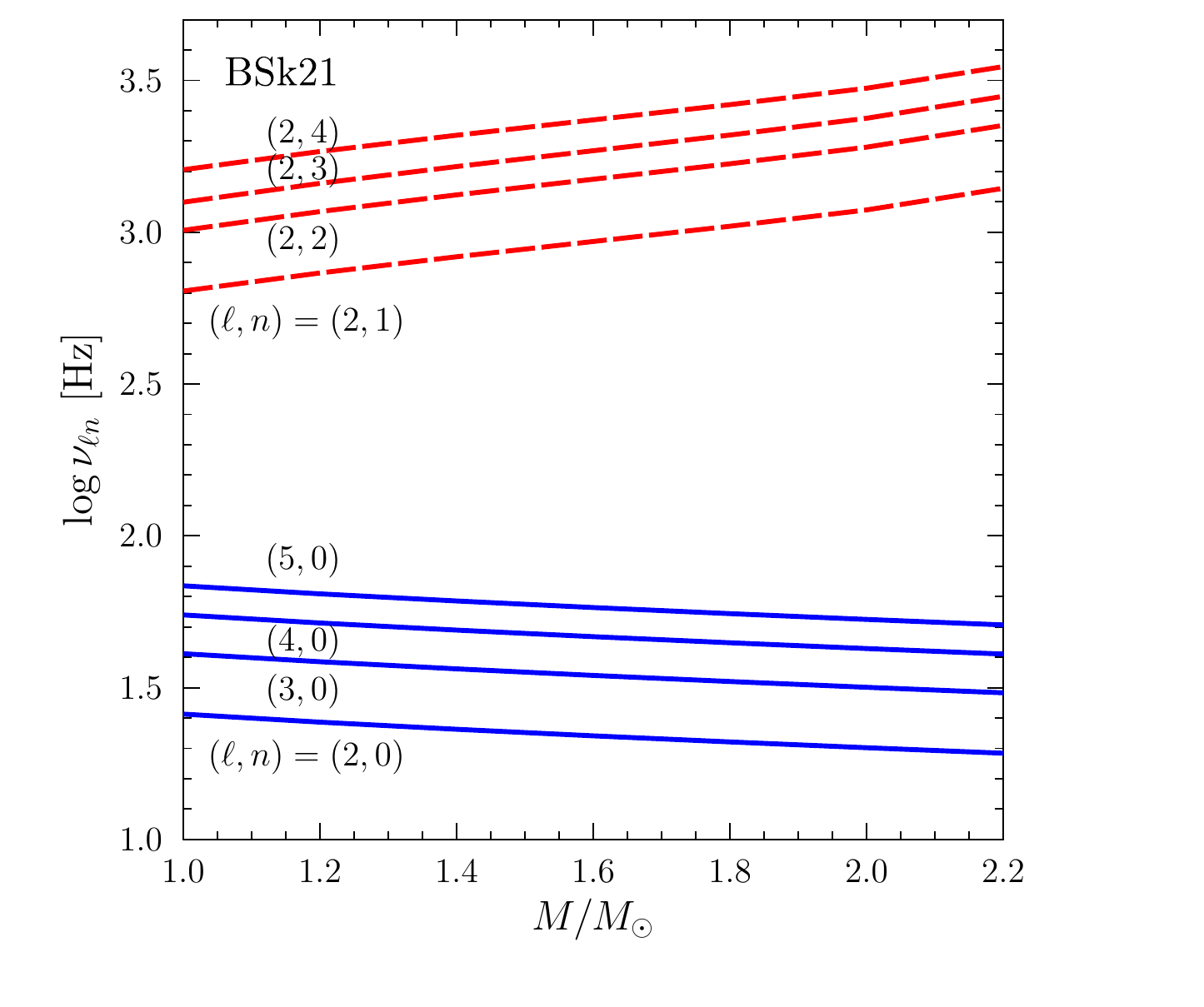}%
	\hspace{5mm}
	\caption{Frequencies $\nu_{\ell n}$  {calculated by the exact method of
		Section \ref{s:torsion} for} some torsional
		vibration modes ($\ell,n$) versus mass $M$ of neutron stars with
		the BSk21 EOS. Four solid (bottom) lines correspond to lowest
		fundamental modes ($n=0$) with $\ell=$2, 3, 4 and 5. Four dashed
		upper lines refer to ordinary modes with lowest $\ell=2$ and
		$n=$1, 2, 3 and 4.
	}
	\label{f:bsk}
\end{figure}

\begin{figure}
	\includegraphics[width=0.5\textwidth]{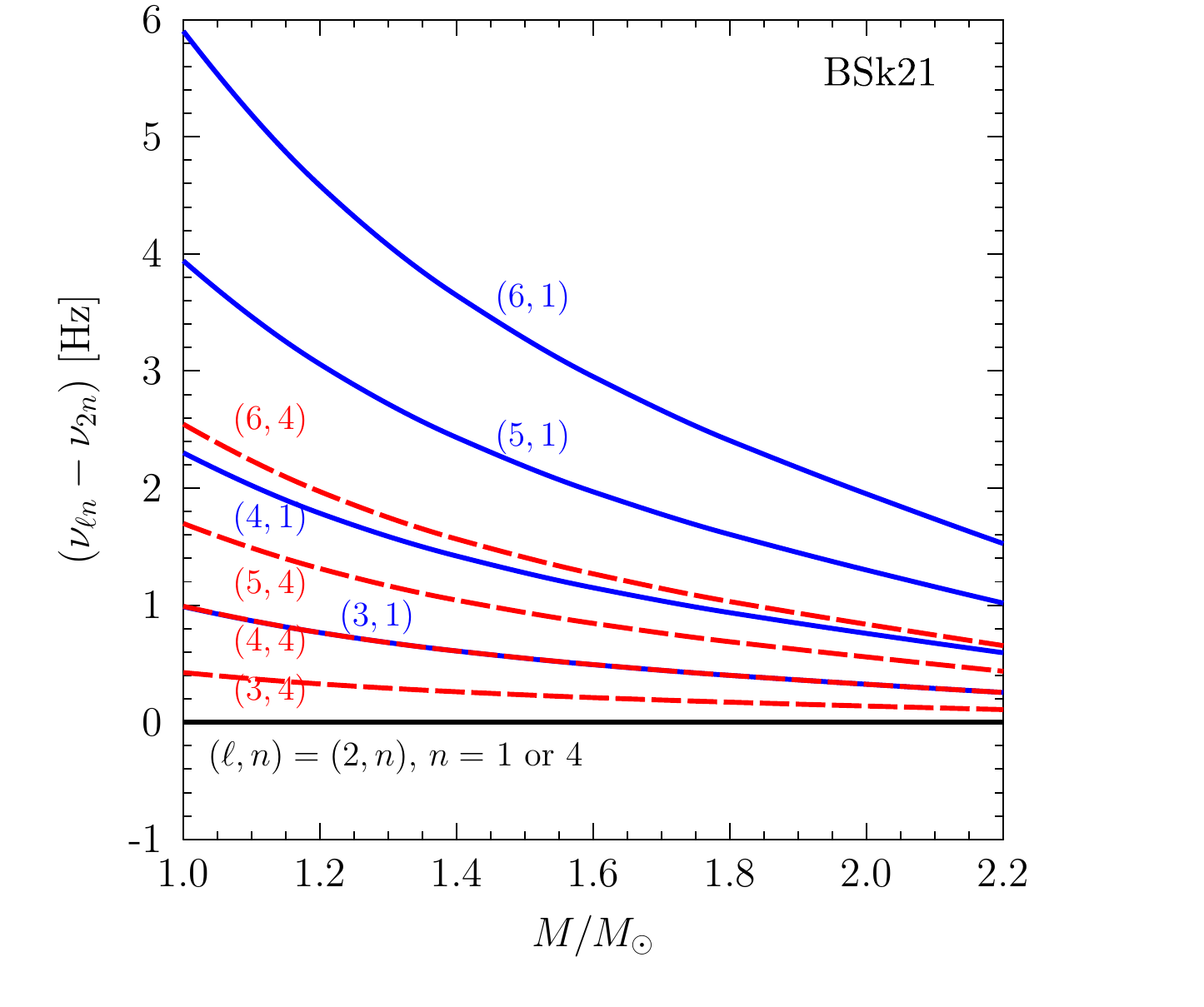}%
	\hspace{5mm}
	\caption{Fine splitting of  {exactly calculated} ordinary vibration frequencies at
		$n=1$ and $n=4$ (solid and dashed lines, respectively) for	
		neutron stars with the BSk21 EOS as a function of $M$.
		Splittings of frequencies $\nu_{\ell n}$ are
		measured from the lowest frequencies $\nu_{21}$ or
		$\nu_{24}$ in corresponding sequences (from the
		lowest horizontal line). For both cases, $n=1$ and
		4, five fine-structure components are shown (with $\ell$
		from 2 to 6). 
	}
	\label{f:bsk-fs}
\end{figure}

The fundamental oscillations ($n=0$) are remarkable. Here, the
centrifugal effect is most important, and the centrifugal
term nearly compensates the shear-wave propagation term in (\ref{e:eqforY}) 
over the inner crust. The shear-momentum transfer in meridional ($\theta$)
direction greatly exceeds the transfer in the radial direction so that
standing waves are typically formed (e.g. \citealt{2012Gabler} ) during propagation of shear perturbations
with velocity $v_{\rm s}$ over meridional directions (over typical  {length-}scales
$\sim \mathrm{\pi}R$). It takes long time and leads to low oscillation frequencies $\nu \sim 20 - 100$ Hz. The crystalline crust appears almost 
non-deformed and only slightly stressed,  {$Y_{\ell 0}(r) \approx Y_{ {\ell 0}}^{ {\rm (a)}} = Y_0$} being a 
good first-order approximation \citep{2020KY}.  {The solutions obtained in this approximation will be labeled with upperscipt (a).}

Then equations (\ref{e:omegaBA})--(\ref{e:omegaB1}) yield
 {$B_{2 \ell 0}^{(\rm a)}=0$}, while  {$A_{ \ell 0}^{(\rm a)}$} and  {$B_{1 \ell 0}^{(\rm a)}$} become independent of $Y_{ { \ell 0}}(r)$, being given by
simple one  {1D} integrals which contain well defined functions.
Then equation (\ref{e:generalnu})  gives  {$\delta \widetilde \nu_{\ell 0}^{\rm (a)}=0$}, and 
the fundamental oscillation frequencies reduce to 
\begin{equation}
    {
      \nu_{\ell 0}^{\rm (a)}=\nu_{20}^{\rm (a)}\,\sqrt{\frac{(\ell-1)(\ell+2)}{4}},\quad
      \nu_{20}^{\rm (a)}=\frac{\delta \widetilde{\nu}_{\ell 0}^{\rm (a)}}{2}=\frac{1}{4\mathrm{\pi}}\,
      \sqrt{\frac{B_{1 20}^{\rm (a)}}{A_{20}^{\rm (a)}}}.
    }
\label{e:nu-ell0}      	
\end{equation}
Therefore, all fundamental frequencies (for a given neutron star model) 
are expressed through the lowest frequency  {$\nu_{20}^{\rm (a)}$}. The latter is easily
calculated using (\ref{e:omegaA}) and (\ref{e:omegaB1}). The first
equation (\ref{e:nu-ell0})  {presents selfsimilarity relation
for fundamental torsional oscillation. It} has been pointed out in a number of 
publications (e.g., \citealt{2007Samuel,2016Gabler}). 
Its more strict derivation was given by \citet{2020KY}. 

Fig. \ref{f:bsk} shows  {exactly calculated} frequencies (solid lines) of four lowest fundamental torsional oscillation modes ($\ell, n$)=(2,0), (3,0), (4,0) and (5,0). The frequencies are calculated for neutron stars built of matter with the BSk21 equation of state (EOS), as discussed in Section \ref{s:bsk}. They are plotted versus neutron star mass $M$.  {The approximately calculated frequencies
(\ref{e:nu-ell0}) turn out to be virtually exact, being equal to the exact ones within estimated numerical errors of calculations ($\sim 0.001$ Hz).}

\subsubsection{Ordinary oscillations and fine splitting}
\label{s:ordinary}

The main differnece of ordinary torsional oscillations ($n>0$) from 
the fundamental ones is that the centrifugal term
in equation (\ref{e:eqforY}) is now much smaller than the shear-wave-propagation 
term. In the first approximation, one can neglect the centrifugal
term,  {which} is equivalent to setting $\ell=1$ in (\ref{e:eqforY}). 
This corresponds to purely radial shear wave oscillations.  {Formally,
the case of $\ell=1$ is well known to be forbidden in the adopted relativistic Cowling approximation (it would violate angular momentum conservation). 
Nevertheless, the solution of (\ref{e:eqforY}) at $\ell=1$ does exist, and gives
a valid approximate solution for radial wave function of ordinary modes,
\begin{equation}
    Y_{\ell n} \approx Y_{\ell n}^{\rm (a)}=Y_{1 n}.
    \label{e:nu-ell1}
\end{equation}
This approximation is supported by the results of \citet{2020KY}.
It is clear that the approximate integrals $A_{\ell n}^{\rm (a)}\equiv {\rm A}_n$, $B_{2 \ell n}^{\rm (a)}\equiv {\rm B}_{2n}$ and $B_{1 \ell n}^{\rm (a)}\equiv {\rm B}_{1n}$, calculated from equations (\ref{e:omegaA})--(\ref{e:omegaB1}), become {\it independent} of $\ell$.
Accordingly, the auxiliary frequencies $\widetilde \nu_{\ell n}^{\rm (a)}\equiv
\widetilde \nu_n$ and $\delta \widetilde \nu_{\ell n}^{\rm (a)}\equiv
\delta \widetilde \nu_n$ are also independent of $\ell$, and the approximate
oscillation frequencies are given by
\begin{equation}
\nu_{\ell n}^{\rm (a)} = \left[  \widetilde{\nu}_n^2
+ (\ell+2)(\ell-1) \, \delta \widetilde{\nu}_n^2  \right]^{1/2} .     
\label{e:generalnua} 
\end{equation}
}

 {As a result, a sequence of oscillation frequencies $\nu_{\ell n}$ 
for a fixed $n$ at different $\ell=2,3,...$ is
determined by two easily calculable 
constants $\widetilde{\nu}_n$ and $\delta \widetilde{\nu}_n$. 
The problem of finding the total spectrum $\nu_{\ell n}$ of
frequencies for ordinary modes with fixed $n$ reduces to determining one a pair of constants, $\widetilde{\nu}_n$ and $\delta \widetilde{\nu}_n$. This is another selflimilarity relation valid for ordinary modes, in addition to
equation (\ref{e:nu-ell0}) for fundamental modes. Actually, the latter equation can also be dersribed by equation (\ref{e:generalnua}) with $\widetilde{\nu}_0=0$, so that both equations have the same nature. They reveal {\it universal} $\ell$-dependence of oscillation frequencies at any given $n$.
}

 {For ordinary modes, in contrast to fundamental ones, the meridional shear momentum transfer
is much weaker than the radial one ($\delta \widetilde \nu_n \ll \widetilde \nu_n$). Typical
pulsation periods are determined by the time of radial  shear wave propagation
through the inner crust (much shorter than for fundamental modes). 
Accordingly, the ordinary torsional pulsation frequencies $\nu_{\ell m} \gtrsim
500$ Hz are much higher than for fundamental ones.
For example, the dashed curves in Fig. \ref{f:bsk} demonstrate exactly
calculated  
frequencies $\nu_{2n}$ at $\ell=2$ and $n=$1, 2, 3 and 4 for stars of different
masses with the BSk21 EOS. In logarithmic scale, the dashed
curves look nearly equidistant, so that ratios of any pair of
frequencies $\nu_{\ell n}$, plotted by dashed lines, are almost independent of $M$. The approximate frequencies turn out to be virtually exact, just as 
for fundamental modes; see
Section \ref{s:n=0}.}

 {Formally, the pulsation spectrum is now given by equation (\ref{e:generalnua}).
In contrast to fundamental modes, $\widetilde{\nu}_n$ is not zero, but it is  
much larger than $\delta \widetilde{\nu}_n$. Actually, $\delta \widetilde{\nu}_n$ at $n>0$ has the same nature as at $n=0$.}

If one fixes $n$ and increases $\ell$, one obtains a sequence of oscillation modes with 
with slightly higher frequencies. This can be treated as {\em a  `fine-structure' splitting}
of basic frequencies $\nu_{2 n}$. The  {splitting} 
occurs for all frequencies plotted in Fig. \ref{f:bsk} by dashed curves.
 {It is} too small to be visible in Fig. \ref{f:bsk} in logarithmic scale.  {It} is demonstrated in Fig.\ \ref{f:bsk-fs} for the two cases of $n=1$ (solid lines) and $n=4$ (dashed lines). In each case, five fine-structure 
shifts of components are shown, with  {$\ell$ varying} from 2 to 6. 
The bottom horizontal line is the basic frequency ($\nu_{21}$ or $\nu_{24}$). Other lines show fine-structure shifts of
higher-$\ell$ components. The fine splitting is seen to be really small.

This smallness is a natural consequence of
the fact that one typically has $\delta \widetilde \nu_n \ll
\widetilde \nu_n$ at $n>0$ under physical conditions in neutron star
crust. Small ratios $\delta \widetilde \nu_n/\widetilde \nu_n
\lesssim 0.01$ are associated with the smallness of shear
velocity with respect to ordinary sound velocity in the crust ($\mu \ll P$). Note also, that at very large $\ell$ the auxiliary
frequencies $\delta \widetilde \nu_n$ and $\widetilde \nu_n$ may
start to depend on $\ell$ and the  {validity of the approximate} 
approach may be broken but it may be not very important for
applications. The estimates show that it may happen at
$\ell \gg 10$.

\section{Neutron stars with BSk21 EOS}
\label{s:bsk}

\subsection{Neutron star models}
\label{s:NSmodels}

Here, by way of illustration, the theoretical consideration
of Section \ref{s:freq} is applied to neutron stars composed
of matter with the BSk21 EOS. Various properties of this
matter have been accurately approximated by analytic expressions
by \citet{BSk2013}. The EOS is unified -- based on the same energy-density functional theory of nuclear interactions in 
the core and the crust. The liquid core consists
of neutrons, protons, electrons and muons, while the crust contains 
spherical atomic nuclei and electrons, as well as quasi-free
neutrons and  {some} admixture of quasi-free protons in the inner crust.
 {For this EOS, the neutron drip density is $\rho_{\rm drip}=4.28 \times 10^{11}$ \gcc, and the crust-core interface occurs at $\rho_{\rm cc}=1.34 \times 10^{14}$ \gcc.}  

The EOS is sufficiently stiff, with the maximum neutron
star mass $M_{\rm max}=2.27\,\msun$. The pulsation frequencies
$\nu_{\ell n}$ and radial wave functions $Y_{\ell n}(r)$ have
been determined  {by using exact theoretical formulation (Section \ref{s:torsion}) with the standard boundary conditions, as in
\citet{2020KY}, and with the standard expression for the shear modulus,
that was first derived by \citet{1990Ogata}. It has been checked that the approximate
frequencies $\nu_{\ell n}^{\rm (a)}$ are virtually exact, as was already pointed out in Section \ref{s:torsion}. Accordingly, hereafter the upperscript (a) will be mostly dropped. However, one should bear in mind that
the virtual exactness may be violated at very large $\ell$; see the end of Section \ref{s:freq}.} 

To be specific, a representative set of models has been chosen with
$M/\msun$=1, 1.2, 1.4,\dots 2.2. In several cases, some models with
intermediate $M$ have been considered to check smooth behaviour of
the results as a function of $M$.

\subsection{Oscillations of the 1.4 $\msun$ star}
\label{s:1.4MsunBsk}

\renewcommand{\arraystretch}{1.2}
\begin{table}
	\caption{Some torsional oscillation parameters for  a $1.4\,\msun$ neutron star with the BSk21 EOS
		($Y_0$ is expressed in radians)}
	\label{tab:1.4Msun}
	\begin{tabular}{c c c }
		\hline 
		$\ell$, $n$	& $\nu_{\ell n}$ [Hz] & $E^{\rm vib}_{\ell n}/Y_0^2$ [erg]  \\
		\hline
		2, 0   & 23.06 &  $9.223 \times 10^{48}$ \\
		2, 1   & 830.1  & $1.899 \times 10^{49}$  \\
		2, 2   & 1327.3  & $3.348 \times 10^{48}$ \\
		2, 3   & 1644.8  & $1.990 \times 10^{48}$ \\
		2, 4   & 2086.7  & $3.898 \times 10^{48}$ \\
		3, 0   & 36.45 &  $3.302 \times 10^{48}$  \\
		3, 1   & 830.7 &  $2.721 \times 10^{49}$  \\
		3, 2   & 1327.8 &  $4.840 \times 10^{48}$ \\
		3, 3   & 1645.2 &  $2.843 \times 10^{48}$ \\
		3, 4   & 2086.9 &  $5.572 \times 10^{48}$ \\
		4, 0   & 48.91 &  $7.729 \times 10^{48}$  \\
		4, 1   & 834.5 &  $3.543 \times 10^{49}$  \\
		4, 2   & 1328.5 &  $6.285 \times 10^{48}$ \\
		4, 3   & 1645.7 &  $3.685 \times 10^{48}$ \\
		4, 4   & 2087.3 &  $7.226 \times 10^{48}$ \\
		\hline
	\end{tabular}	 
\end{table}
\renewcommand{\arraystretch}{1.0} 

The model with $M=1.4\,\msun$ has been chosen as basic. In this
case the stellar radius is $R=12.60$ km, and the radius of the
crust-core interface is $R_{\rm cc}=11.55$ km. The central density
of the star is $\rho_{\rm c}=7.30 \times 10^{14}$ \gcc.

Table \ref{tab:1.4Msun} presents 15 torsional oscillation
frequencies $\nu_{\ell n}$ of the  {basic star} with $\ell=$ 2, 3 and
4 and $n=$ 0, 1,\dots 4 (including fundamental 
and ordinary oscillation oscillation modes, 
particularly, fine splitting). 
While calculating $\nu_{\ell n}$, the auxiliary frequencies
$\widetilde{\nu}_{n}$ and  $\delta \widetilde{\nu}_{n}$ have
been obtained and their independence of $\ell$
has been checked.

The last column of Table \ref{tab:1.4Msun} presents vibrational
energies  {$E^{\rm vib}_{\ell n}$} of the same oscillations models  computed
from equation (\ref{e:energy}). In the adopted linear approximation,
all energies  {$E^{\rm vib}_{\ell n}$} are proportional to the squared angular
vibration amplitude, $Y_0^2$, of the surface layer. 
The presented expressions are valid at
$Y_0 \ll 1$ (see \citealt{2020KY}).

\subsection{Dependence on neutron star mass}
\label{s:BSkAllMasses}

\renewcommand{\arraystretch}{1.2}
\begin{table*}
	\caption{Constants $\nu_{02}$, $\widetilde \nu_{n}$ and $\delta \widetilde\nu_{n}$, which determine all vibration frequencies
		of fundamental ($\ell, 0$) and ordinary ($\ell, n$) torsional vibration modes with $n \leq 4$,
		for neutron stars with the BSk21
		EOS and $M/\msun=1,\,1.2,\,1.4,\,1.6,\,1.8,\,2$ and 2.2}
	\label{tab:main1}
	\begin{tabular}{ c c c c c c c c c c c }
		\hline 
		$M$ ($\msun$)     & $R$ (km)& $\nu_{20}$ (Hz)& $\widetilde\nu_1$ (Hz)&  $\delta \widetilde\nu_1  $  (Hz)& $\widetilde\nu_2$ (Hz)&  $\delta \widetilde\nu_2  $ (Hz)& $\widetilde\nu_3$ (Hz)&  $\delta \widetilde\nu_3  $ (Hz) & $\widetilde\nu_4$ (Hz)&  $\delta \widetilde\nu_4  $ (Hz)\\  
		\hline
		1.0 &  12.48 & 25.87 & 639.4  & 14.52 & 1014  & 16.56 & 1255& 15.09 
		& 1606 & 15.08  		 		 
		\\ 		
		1.2 &  12.56 & 24.34 & 733.5  & 13.69 & 1169  & 15.66 & 1448& 14.49 
		& 1844 & 14.21  		 		 
		\\ 
		1.4 &  12.60 & 23.06 & 829.7  & 12.98 & 1327  & 14.87 & 1645& 13.94 
		& 2086 & 13.47		 
		\\   
		1.6 &  12.59 & 21.94 & 931.7  & 12.37 & 1494  & 14.17 & 1855& 13.44 
		& 2344 & 12.85	            	 
		\\ 		    
		1.8 &  12.50 & 20.95 & 1046  & 11.83 & 1680  & 13.54 & 2089& 12.97 
		& 2631 & 12.28
		\\  		    
		2.0 &  12.30 & 20.05 & 1184  & 11.33 & 1906  & 12.97 & 2373& 12.53 
		& 2980 & 11.77
		\\	
		2.2 &  11.81 & 19.23 & 1395  & 10.88 & 2248  & 12.44 & 2803& 12.14 
		& 3511 & 11.30
		\\	       	    
		\hline		
	\end{tabular}	
\end{table*}
\renewcommand{\arraystretch}{1.0} 

\renewcommand{\arraystretch}{1.2}
\begin{table}
	\caption{Fit parameters $f_n$, $\alpha_n$, $\beta_n$,
		$\delta f_n$, $\delta \alpha_n$, and $\delta \beta_n$ in
		equations (\ref{e:Fitnuln}) and (\ref{e:Fitdnuln}) for neutron stars with the BSk21 EOS}
	\label{tab:main2}
	\begin{tabular}{c c c c c c c}
		\hline 
		$n$ & 
		$\delta f_n$ [Hz] &  $\delta \alpha_n$ & $\delta \beta_n$ &
		$ f_n$ [Hz] &   $\alpha_n$ & $ \beta_n$ \\ 
		\hline
		0	&	44.59 &  2.411 &  $-$1.968 &	0 &  -- &  -- \\
		1	&	24.61 &  2.166 & --1.801 & 1171.7 &  $-$1.508 & 1.326 \\
		2	&	27.35 &  1.750 & --1.340 &	1821.5 & $-$1.342 & 1.163 \\
		3	&	22.04  & 0.2639 & --0.2714 &	2245 &  $-$1.316 &  1.166 \\
		4	&	25.60  & 2.198 & --1.843  &	2936 & $-$1.488 & 1.306 \\
		\hline
	\end{tabular}	
\end{table}
\renewcommand{\arraystretch}{1.0} 

\begin{figure}
	\includegraphics[width=0.5\textwidth]{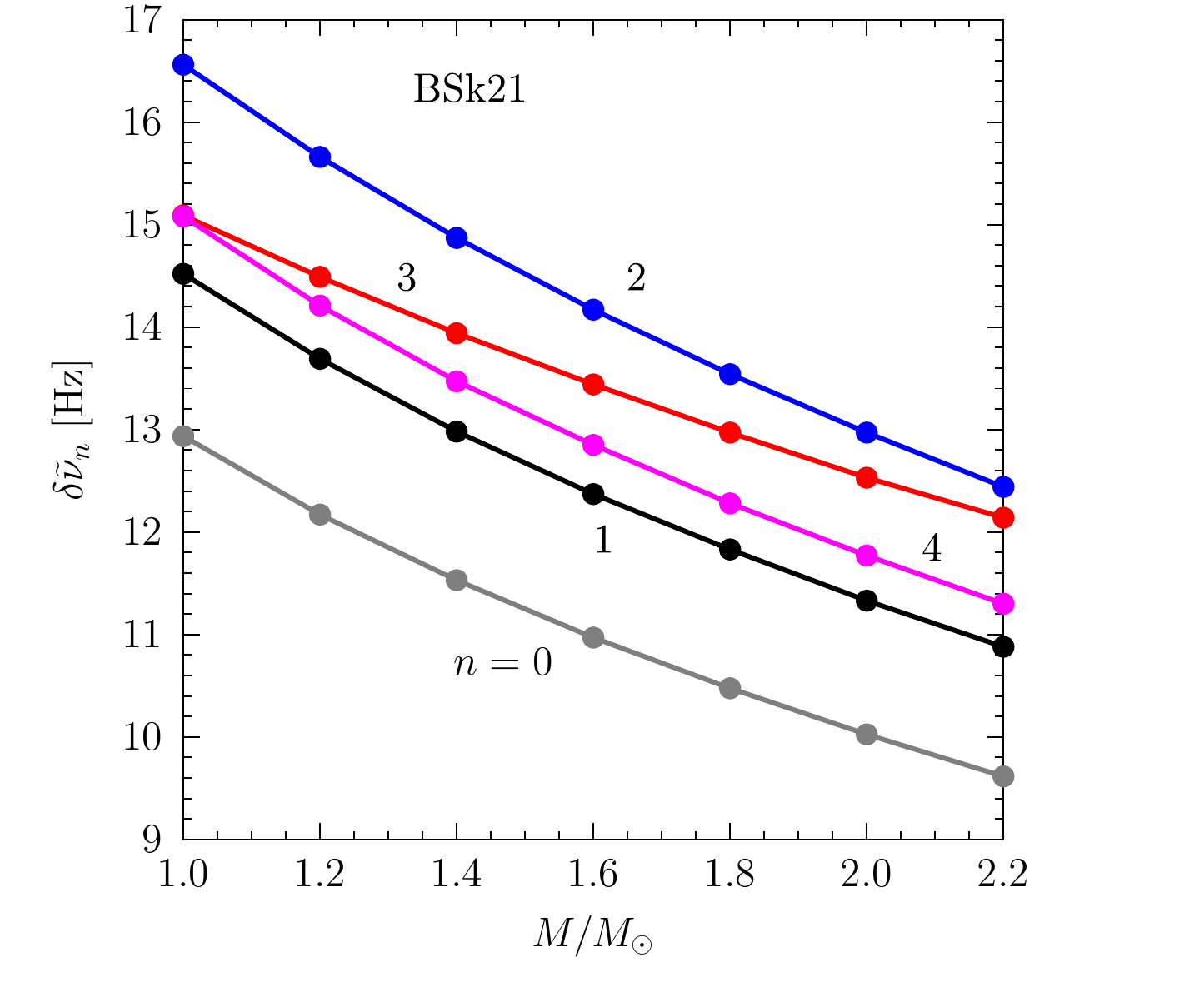}%
	\hspace{5mm}
	\caption{Auxiliary frequencies $\delta \widetilde{\nu}_n$
		in equation (\ref{e:generalnu}) for neutron stars 
		(with the BSk21 EOS) of different masses  
	    at $n=$0, 1,\ldots 4.
	    Filled dots show the calculated values (Table \ref{tab:main1});
	    lines are analytic approximations (equation (\ref{e:Fitdnuln})).
	}
	\label{f:dnu_n(m)}
\end{figure}

\begin{figure}
	\includegraphics[width=0.5\textwidth]{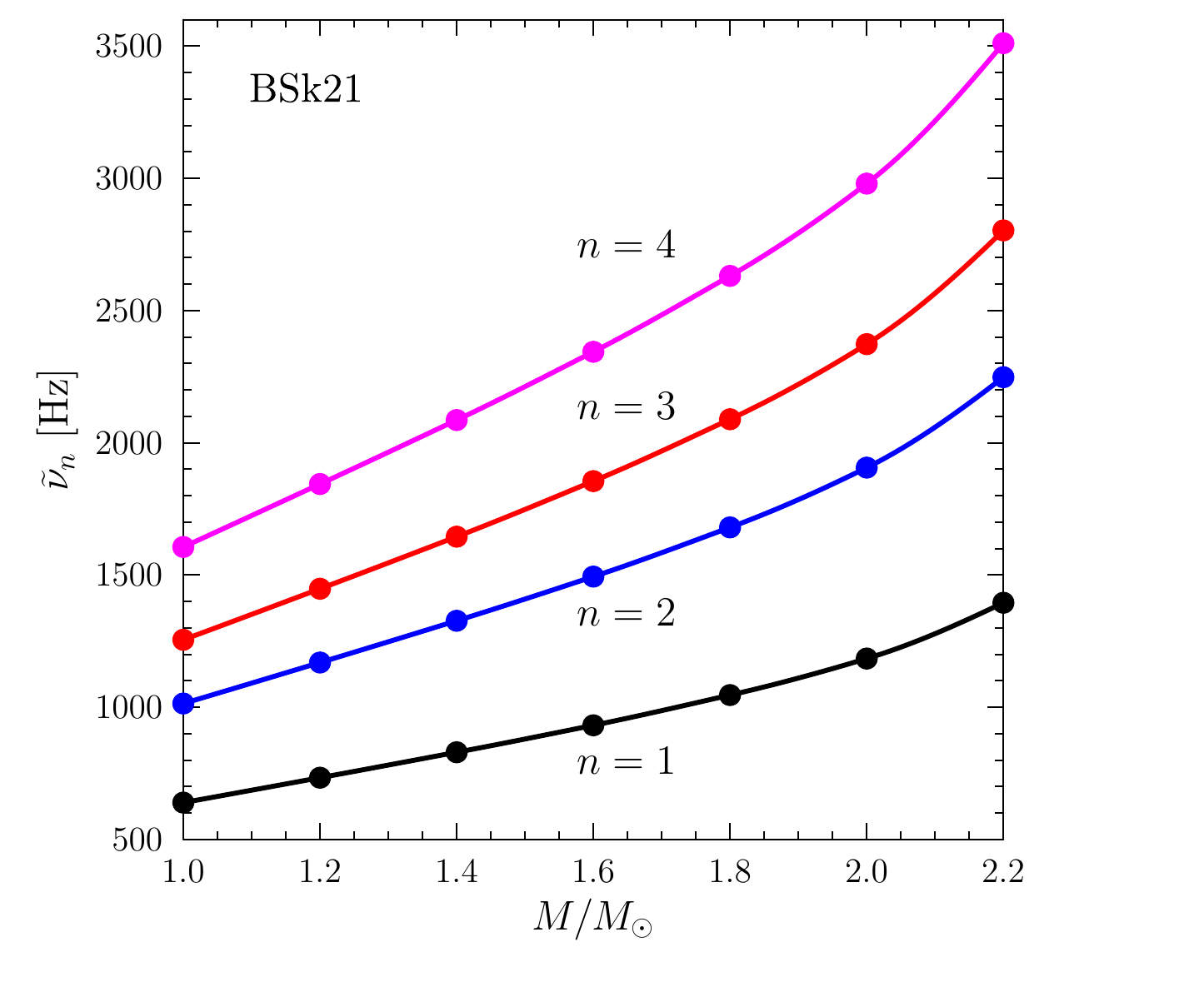}%
	\hspace{5mm}
	\caption{The same as in Fig.\ \ref{f:dnu_n(m)} but 
		for the auxiliary frequencies $\widetilde{\nu}_n$
		at $n=$ 1,\ldots 4 (with $\widetilde{\nu}_0=0$). Lines show analytic approximations
		(\ref{e:Fitnuln}).		
	}
	\label{f:nu_n(m)}
\end{figure}

At the next step the consideration of 
Section \ref{s:1.4MsunBsk} is extended 
to neutron stars of
different masses. To this aim, the frequencies
of the same 15 vibration modes as in
Table \ref{tab:1.4Msun} have been computed for seven values of 
$M/\msun=$1, 1.2,\ldots,2.2. For any $M$,
the auxiliary frequencies $\delta \widetilde{\nu}_{n}$ and   $\widetilde{\nu}_{n}$ have also been found at $n \leq 4$.
These results are presented in Table \ref{tab:main1} and
plotted in Figs.\ \ref{f:dnu_n(m)} and \ref{f:nu_n(m)}.
The second column of Table \ref{tab:main1} lists the  radii $R$ of
neutron star models.

\renewcommand{\arraystretch}{1.2}
\begin{table}
	\caption{Fit parameters $a_n$, $c_n$ and $\delta c_n$,
	 in	equation (\ref{e:FitA})  for neutron stars with the BSk21 EOS}
	\label{tab:energy}
	\centering
	\begin{tabular}{c c c c }
		\hline 
		$n$ & $a_n$  &  $c_n$ & $\delta c_n$  \\ 
		\hline
		0	& 0.924 &  --2.339 &  2.110  \\
		1	& 0.0133 & --2.531 &  2.307   \\
		2	& 0.000844 & $-$2.680 & 2.418  \\
		3	& 0.000620  &  $-$1.162 & 0.933  \\
		4	& 0.000623  &  $-$1.762 &1.635   \\
		\hline
	\end{tabular}	
\end{table}
\renewcommand{\arraystretch}{1.0} 

To simplify using these results, all auxiliary
frequencies have been fitted by analytic functions of
neutron star mass and radius, 
\begin{equation}
	\delta \widetilde{\nu}_{ n}
	= 
	\frac{\sqrt{1-\xg } \, \delta f_n}{R_{10} 
		\sqrt{1+ \delta \alpha_n \xg+ \delta \beta_n x_{\rm g}^2}},
	\label{e:Fitdnuln}	
\end{equation}
\begin{equation}
	\widetilde \nu_{ n }
	=\frac{M_1}{R_{10}^2} f_{n}
	\sqrt{1+\alpha_n  \xg + \beta_n x_{\rm g}^2},
	\label{e:Fitnuln}
\end{equation}
where  {$\xg \sim 0.3$ is the neutron star
compactness parameter that is defined here as $\xg=r_{\rm g}/R= {2GM/(Rc^2)}$}; $R_{10}=R/10$ km,
$M_1= M/\msun$; $f_n$, $\alpha_n$ and $\beta_n$ are
fit parameters for $\widetilde \nu_n$;  $\delta f_n$,
$\delta  \alpha_n$ and $\delta \beta_n$ are those
for $\delta \widetilde \nu_n$.

The structure of  {these} fits
reflects the nature of slow and faster wave propagations
in fundamental and ordinary torsional oscillations
(Section \ref{s:n=0andn>0}). A characteristic time-scale
of the slow meridional heat transport is $\delta \tau \sim v_{\rm s*}R/\sqrt{1-\xg}$,
where $v_{\rm s*}$ is a typical shear velocity (\ref{e:vsound}) 
in the inner crust, and $1/ \sqrt{1-\xg}$ is the correction
factor for a distant observer. Accordingly, 
equation (\ref{e:Fitdnuln}) contains the factor  {$1/\delta \tau
\propto 1/R$}.
Analogously, characteristic faster radial propagation time scale 
(with typical shear-sound velocity  {$v_{\rm s*}$})
is $\tau \sim v_{\rm s*} \Delta R$, where $\Delta R$ is a proper
depth of the inner crust. The latter can be estimated from the
equation of local hydrostatic equilibrium, 
$\Delta R \sim P_*/(g_{\rm s} \rho_*)$, where $P_*$ and
$\rho_*$ are typical pressure and density in the inner crust and 
$g_{\rm s} =GM/(R^2 \sqrt{1-\xg})$ is the surface gravitational
acceleration. As a result, the proper scaling factor in
equation (\ref{e:Fitnuln}) is $M/R^2$ (the factor $\sqrt{1-\xg}$ cancels out while transforming to the
reference frame of a distant observer). Note that scaling relations similar to (\ref{e:Fitdnuln}) and (\ref{e:Fitnuln}) (although less refined) have been discussed in many publications 
(e.g. \citealt{2007Samuel,
2012Gabler}).

Naturally, introducing the above scaling factors in 
(\ref{e:Fitdnuln}) and (\ref{e:Fitnuln}) is not enough to
ensure high fit accuracy. One needs  {some}
additional tuning. It is done by inserting 
extra square-root factors containing correcting
fit parameters $\alpha_n$, $\beta_n$, $\delta \alpha_n$
and $\delta \beta_n$. After that the auxiliary frequencies from
Table \ref{tab:main1} have been fitted by
equations (\ref{e:Fitdnuln}) and (\ref{e:Fitnuln}). 
The resulting fit parameters are listed in Table \ref{tab:main2}.
These fits are very accurate, with maximum relative deviations
of the fitted $\widetilde \nu_{ n }$ and $\delta \widetilde \nu_{ n }$
values from those in Table \ref{tab:main1} smaller than 0.003.

The fits (\ref{e:Fitdnuln}) and (\ref{e:Fitnuln}), 
combined with Table 
\ref{tab:main2}, allow one to calculate torsional pulsation
frequencies with $n \leq 4$ at
any $\ell$ and at  $M/\msun$ from 1 to 2.2. It seems to be the most compact representation
of torsional oscillation frequencies $\nu_{\ell n}$.

Now one can introduce relative deviations 
$\delta_{\ell n}= |\nu_{\ell n}^{\rm fit}/\nu_{\ell n}^{\rm calc}-1|$
of the fitted frequencies from the calculated  ones. 
The root-mean-square (rms) deviation over all 105 frequencies is
$\delta_{\rm rms} \approx 0.0021$, and
the maximum deviation $\delta_{\rm max} \approx 0.008$
occurs at $\ell=4$, $n=1$ and $M=1 \msun$. For the problem
of study, such a fit accuracy seems too good. Although
the fit is done for seven values of $M$, it has been checked that
it remains equally accurate for intermediate values.

It would be no problem to extend the results for $n>4$. One often 
assumes [e.g. equation (21) in \citealt{2007Samuel}] that with increasing $n$ the frequency $\nu_{\ell n} \approx
\widetilde \nu_{2n}$ scales proportionally to $n$. This assumption is based on analogy
with  {1D} oscillations in  a uniform slab. It 
is approximately confirmed by calculations of $\nu_{2n}$ at
$n=$ 2, 3 and 4 by \citet{2007Sotani} for neutron stars with
several EOSs and masses (see their table 4 and Section \ref{s:previous} below).
However,
this scaling is not the law of nature and can be violated, as
seen from Fig.\  \ref{f:nu_n(m)}. As for the auxiliary frequency
$\delta \widetilde \nu_{ n }$, it is rather small and unimportant
in the total frequency $\nu_{\ell n}$. It is expected that replacing
$\delta \widetilde \nu_{ n }$ by  $\delta \widetilde \nu_{4}$ at
$n>4$ would be a reasonable approximation for calculating
$\nu_{\ell n}$.

In addition, the quantity  {$A_{\ell n} \approx A_n^{\rm (a)}$ }
given by equation (\ref{e:omegaA}) has been
calculated.  {Here we again use the uppersript (a) to distinguish between approvimate and exact quantites. $A_n^{\rm (a)}$}   is needed to determine the approximate
vibrational energy  {$E^{\rm vib(a)}_{\ell n}$} from equation (\ref{e:energy}). Calculations have been
done at $n \leq 4$ for the same seven values of $M$.
The results are fitted by the expression 
\begin{equation}
	 {A_n^{\rm (a)}}=10^{42} \, \frac{Y_0^2 R_{10}^6\, a_n}{M_1 \,
		(1+c_n x_g + \delta c_n x_g^2)} \;\; \,{\rm g \, cm^2},
	\label{e:FitA}	
\end{equation}
where $a_n$, $c_n$ and $\delta c_n$ are dimensionless fit
parameters listed in Table \ref{tab:energy}. Exact  
values of  {$E_{\ell n}^{\rm vib}$} have been computed 
from equations (\ref{e:omegaA}) and (\ref{e:energy}) for the same
105 oscillation modes ($\ell,n$). The rms
relative error of the fits is $\approx 0.01$, while the maximum
error $\approx 0.024$ occurs at $\ell=4$, $n=0$ and
$M=1.2 \, \msun$.

\subsection{Torsion oscillations in flat space-time crust}
\label{s:flatspace crust}

\begin{figure}
	\includegraphics[width=0.5\textwidth]{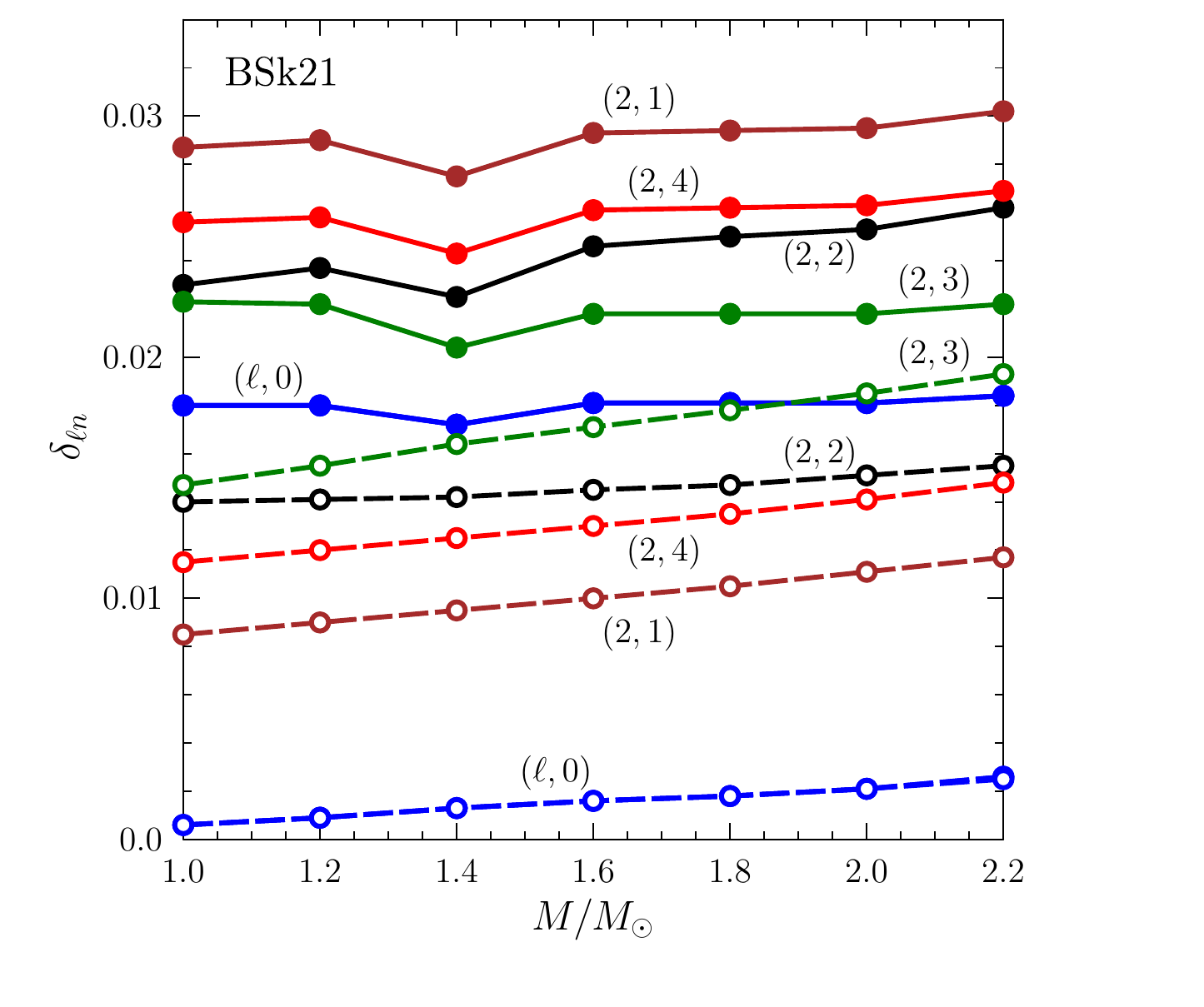}%
	\hspace{5mm}
	\caption{Absolute values of relative deviations $\delta_{\ell n}$ of 
		some vibration frequencies $\nu_{\ell n}^{\rm fs}$, calculated for
		seven values of $M/\msun$=1, 1,2,\ldots 2.2 in the
		flat space-time approximation, from those, calculated in 
		full General Relativity.
		Filled and open dots ( {solid and dashed lines}) refer, respectively, to  $x_{g*}=x_{\rm s}$ and $x_{g*}=x_{\rm cc}$. The vibration modes are ($\ell,n$)=
		(2,0), (3.0), (4,0), (2,1), (2,2), (2,3) and (2,4).  At $n=0$ the
		deviations $\delta_{\ell n}$ are almost independent of $\ell$
	     {and merge in one solid or dashed curve}. 	
	}
	\label{f:err1}
\end{figure}

So far all calculations have been done in full General Relativity. Here
we check if General Relativity is really needed?

A neutron star crust is usually thin (it thickness is $\sim 1$ km) and 
contains about $\sim 1$ per cent of the stellar mass. Naturally,
the metric (\ref{e:metric}) in the crust is expected to be close
to the Schwarzschild metric, with the metric functions $\Lambda(r)$ and
$\Phi(r)$ close to those given by equation (\ref{e:outsideNS}).

 {Let} us try even simpler approach,
with constant $\Lambda$ and $\Phi$  throughout the crust,
\begin{equation}
\Lambda =  -\Phi= -{\textstyle \frac{1}{2}} \ln(1-x_{\rm g*}),
\label{e:FlatMat}   
\end{equation}
where $x_{\rm g*}=2Gm(r_*)/(c^2 r_*)$, and $r_{\rm g*}$ is some
fiducial radial coordinate  {$r=r_*$} in the crust. This makes the space-time (\ref{e:metric}) in the crust artificially flat (although different
from the asymptotically flat space-time of a distant observer). 

We have solved equation (\ref{e:eqforY}) in this approximation for
neutron stars with the BSk21 EOS at the same 7 values of
$M/\msun$ from 1 to 2.2, as in Section \ref{s:BSkAllMasses}, and 
found flat-space (fs) vibration frequencies
$\nu_{\ell n}^{\rm fs}$ for the same values of $\ell$ and $n$ as
in Table \ref{tab:1.4Msun} at several values of $r_*$. For all 105 
 {oscillation modes} at any $r_*$ we have tried,
relative deviations $\delta_{\ell n}=
|\nu_{\ell n}^{\rm fs}/\nu_{\ell n}-1|$ of  {$\nu_{\ell n}^{\rm fs}$
from  exact frequencies $\nu_{\ell n}$ (determined in full General Relativity)
do not exceed a few per cent.}

The results for two cases of (I) $r_*=R$, $x_{\rm g*}=x_{\rm s}$
($x_{\rm g*}$ is the same as at the surface) and (II)
$r_*=R_{\rm cc}$, $x_{\rm g*}=x_{\rm cc}$
(at the crust-core interface) are plotted in Fig.\
\ref{f:err1}. Filled and open dots refer to cases I and
II, respectively. In case I the rms deviation is $\approx 0.024$,
and the maximum deviation is $\approx 0.03$. In case II one has $\delta_{\rm rms}\approx 0.013$ and 
$\delta_{\rm max} \approx 0.02$. 

If one does not need a very high accuracy,
the flat space-time approximation with any $r_*$ from $R_{\rm cc}$
to $R$ is reasonably good. Higher accuracy at $r_*=R_{\rm cc}$
(case I) is quite understandable -- torsional oscillations
are mainly formed in the inner crust. Naturally, the Newtonian approximation is accurate because torsional oscillations are
fully confined in a thin and light crust. For oscillations of other types which
penetrate the neutron star core, the agreement
is much worse (as can be easily deduced from the results of \citealt{2002YL}).

\section{Previous work}
\label{s:previous}

\subsection{Different EOSs}
\label{s:EOS}

\renewcommand{\arraystretch}{1.2}
\begin{table}
	\caption{Quantities  {$\nu_{20}$, $\widetilde \nu_{1}$ 
		and $\delta \widetilde \nu_{1}$} which determine all
		vibration frequencies of fundamental ($0,\ell$) and ordinary ($1,\ell$) torsional vibrations 
		of neutron stars with the 
		EOSs, masses and radia considered by
		\citet{2007Sotani}; see the text for details}
	\label{tab:sotani}
	\begin{tabular}{c c c c c c}
		\hline 
		EoS & $M$      & $R$ & $\nu_{20}$ & $\widetilde \nu_1$&  
		$\delta \widetilde\nu_1  $ \\ 
		 & ($\msun$) & (km) & (Hz) & (Hz) & (Hz) \\ 
		\hline
		A+DH& 1.4 &  9.49 & 28.50 & 1206  & 15.42 \\     
		A+DH &1.6 &8.95 &  27.20  & 1531 & 14.80  \\  
		WFF3+DH &1.4 &10.82& 26.33  & 941.9  & 14.36 \\  
		WFF3+DH &1.6 &10.61& 25.23 & 1101 & 13.76 \\ 
		WFF3+DH &1.8 &10.03& 24.29  & 1367 & 13.17 \\ 
		APR+DH &1.4 &12.10& 24.60  & 760.8 & 13.37 \\ 
		APR+DH &1.6 &12.09& 23.38  & 859.8 & 12.71 \\ 
		APR+DH &1.8 &12.03& 22.28  & 965.2 & 12.17 \\ 
		APR+DH &2.0 &11.91& 21.27  & 1083 & 11.56 \\ 
		APR+DH &2.2 &11.65& 20.18  & 1238 & 11.00 \\ 
		L+DH &1.4 &14.66 &  21.55   & 529.7 & 11.68 \\
		L+DH &1.6 &14.78 & 20.58  & 586.0 & 11.15  \\ 
		L+DH &1.8 &14.83 & 19.67   & 647.7 & 10.71 \\ 
		L+DH &2.0 &14.82 & 18.93  & 712.2 & 10.26 \\ 
		L+DH &2.2 &14.73 & 18.19   & 787.9 & 9.866 \\ 
		L+DH &2.4 &14.54 & 17.54   & 874.0 & 9.564 \\ 
		L+DH &2.6 &14.13 & 16.96   & 995.0 & 9.173 \\ 
		A+NV &1.4 &9.48  & 28.71   & 950.5 & 15.58 \\ 
		A+NV &1.6 &8.94  & 27.38   & 1190 & 14.91 \\
		WFF3+NV &1.4 &10.82& 26.69  & 740.2 & 14.46 \\
		WFF3+NV &1.6 &10.61& 25.41  & 865.4 & 13.80 \\
		WFF3+NV &1.8 &10.03& 24.38  & 1069 & 13.28 \\
		APR+NV &1.4 &11.93& 25.16  & 615.8 & 13.59 \\
		APR+NV &1.6 &11.95& 23.81  & 688.0 & 12.87 \\
		APR+NV &1.8 &11.92& 22.59  & 769.0 & 12.25 \\
		APR+NV &2.0 &11.82& 21.43  & 858.0  & 11.63 \\
		APR+NV &2.2 &11.59& 20.32  &  974.1 & 11.07 \\
		L+NV &1.4 &13.58&  23.17   & 483.0 & 12.46 \\
		L+NV &1.6 &13.82&  21.82  & 524.0 & 11.76  \\
		L+NV &1.8 &14.00&  20.68   & 567.2 & 11.19 \\
		L+NV &2.0 &14.09&  19.68   & 614.5 & 10.68 \\
		L+NV &2.2 &14.11&  18.78   & 666.8 & 10.19 \\ 
		L+NV &2.4 &14.02&  17.96  & 728.8 & 9.780 \\
		L+NV &2.6 &13.68&  17.20  & 823.9 & 9.361 \\
		\hline
	\end{tabular}	
\end{table}
\renewcommand{\arraystretch}{1.0} 

Since the torsional oscillations have been studied in many
publications, it is instructive to analyse previous results
using the formalism of Sections \ref{s:freq} and \ref{s:bsk}.

In particular, an extensive calculation of torsional
oscillation frequencies of non-magnetic spherical 
neutron stars was performed by \citet{2007Sotani}. The
authors presented detailed tables of oscillation frequencies
of stars with different EOSs and 
masses ($M \geq 1.4\,\msun$). They considered fundamental modes $\nu_{\ell 0}$ 
with $\ell$= 2, 3,\dots 10 (their table 2), ordinary
modes with one radial node, $\nu_{\ell 1}$, with the same
$\ell$ (table 3), and lowest-$\ell$ modes, $\nu_{2n}$, with
$n=2$, 3 and 4 (table 4). 

\citet{2007Sotani} took four EOSs in the neutron star
core and two EOSs in the crust. The  {EOSs} in the core were labeled as A (EOS A suggested
by \citealt{1971P}, with $M_{\rm max}\approx 1.65\, \msun$),
WFF3 (\citealt{1988WFF}, $M_{\rm max}\approx 1.8\, \msun$),
APR (\citealt{1998APR}, $M_{\rm max}\approx 2.35\, \msun$)
and L (\citealt{1975PS}, $M_{\rm max}\approx 2.7\, \msun$). 
The EOSs in the crust  {were} labeled as DH and NV; they were constructed 
by \citet{DH2001} and \citet{1973NV}, respectively.
Their use yields almost the same $M_{\rm max}$ for
a fixed EOS in the core.

Recent progress allowed one to accurately measure
(constrain) masses of heavy neutron stars in
binary systems. In particular,
\citet{2013Antoniadis} and \citet{2021Fonseca} reported
the masses of  {two} millisecond pulsars in compact binaries with
white dwarfs ($M=2.01 \pm 0.04\,\msun$ for the PSR J0348+0432,
and $M=2.08 \pm 0.07\,\msun$ for the PSR J0740+6620). 
\citet{2022Romani} measured the mass  
$M=2.35 \pm 0.17\,\sun$ of the 
millisecond ``black widow" pulsar J0952--0607 (in pair with a low-mass
companion). All deviations are given at 1$\sigma$ level.

These  {observations indicate}  that the A and WFF3 EOSs 
are outdated ( {too soft to support} most massive pulsars). However, following
\citet{2007Sotani}, we include them in our analysis in order  {to}
broad the bank of theoretical vibration
frequencies.

\renewcommand{\arraystretch}{1.2}
\begin{table*}
	\caption{Fit parameters $f_n$, $\alpha_n$, $\beta_n$,
		$\delta f_n$, $\delta \alpha_n$, and $\delta \beta_n$ 
		at $n=0$ and 1 in
		equations (\ref{e:Fitnuln}) and (\ref{e:Fitdnuln}) for neutron stars with the BSk21 EOS; see the text for details}
	\label{tab:sotani1}
	\begin{tabular}{c c c c c c c  c c c}
		\hline 
		EOS & 
		$\delta f_0$ [Hz] &  $\delta \alpha_0$ & $\delta \beta_0$ &
		$\delta f_1$ [Hz] &  $\delta \alpha_1$ & $\delta \beta_1$ &
		$ f_1$ [Hz] &   $\alpha_1$ & $ \beta_1$ \\ 
		\hline
		BSk21	&	44.59 &  2.411 &  $-$1.968 &  
		24.61 & 2.166 & --1.801 & 1171.7 &  $-$1.508 & 1.326 \\
		APR+DH & 42.43 & 1.322 & $-$0.958 & 22.36 & 0.978 & --0.6743 & 
		919.7 &--1.015 &0.822 \\ 
		APR+NV &  45.60 & 1.972 & $-$1.459  & 24.68 & 2.081 & --1.719
		& 748.9& --1.218  &   0.997    \\
		L+DH &  43.89 &  1.781 & $-$1.477  & 22.90 & 1.262 &--0.890  &  
		925.3  &--1.066   & 0.887  \\
		L+NV &   45.50 & 1.922 & $-$1.380 & 23.63 & 1.531 & --1.170 
		& 748.7  & --1.213  & 0.978  \\   
		\hline
	\end{tabular}	
	\\
\end{table*}
\renewcommand{\arraystretch}{1.0} 

\begin{figure}
	\includegraphics[width=0.5\textwidth]{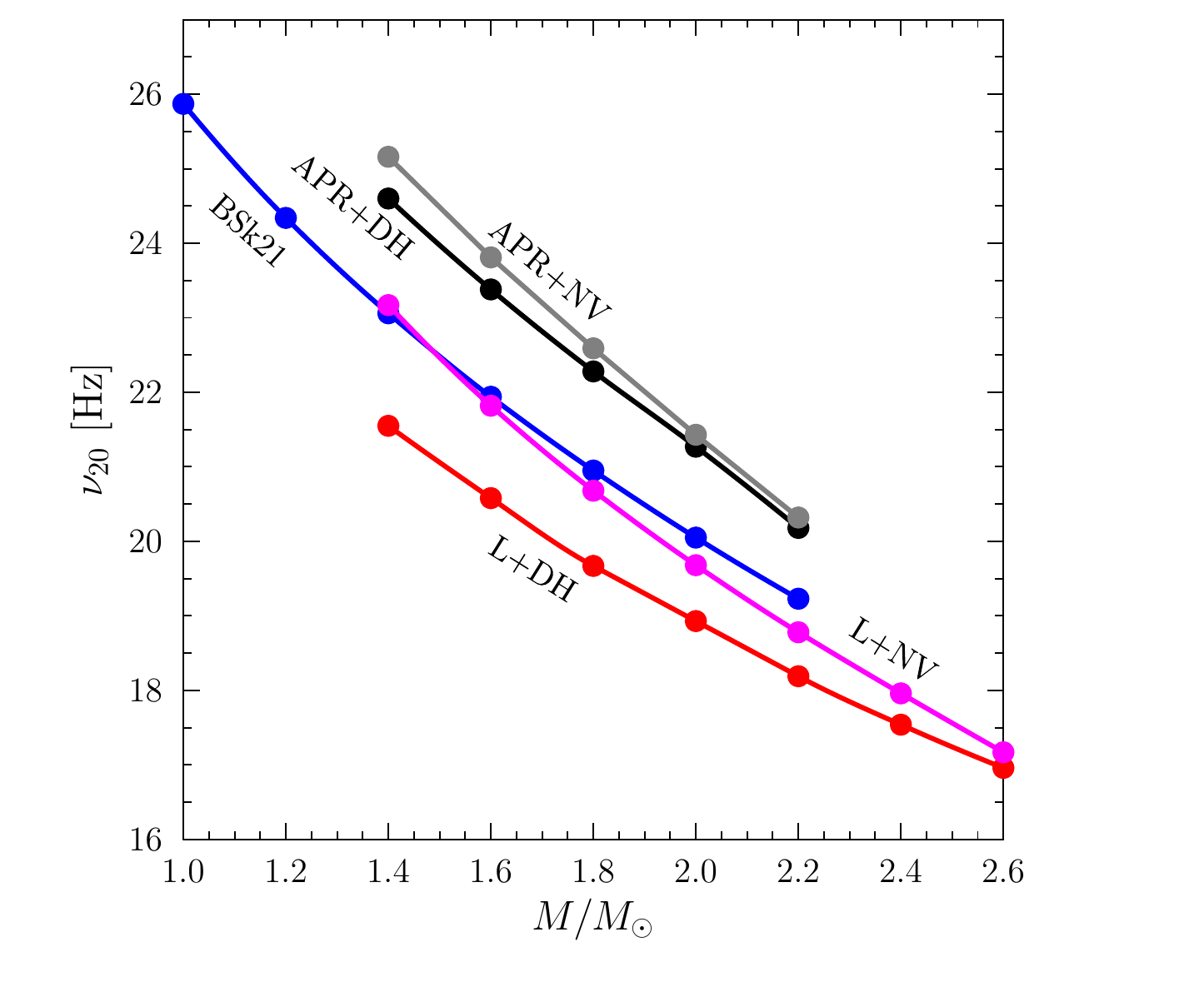}%
	\hspace{5mm}
	\caption{Lowest fundamental oscillation frequency for neutron stars
		as a function of $M$. The stars are composed of
		different EOSs: BSk21, APR+NV,
		APR+DH, L+NV and L+DH. See the text for details. 
	}
	\label{f:nu20(m)}
\end{figure}

\begin{figure}
	\includegraphics[width=0.5\textwidth]{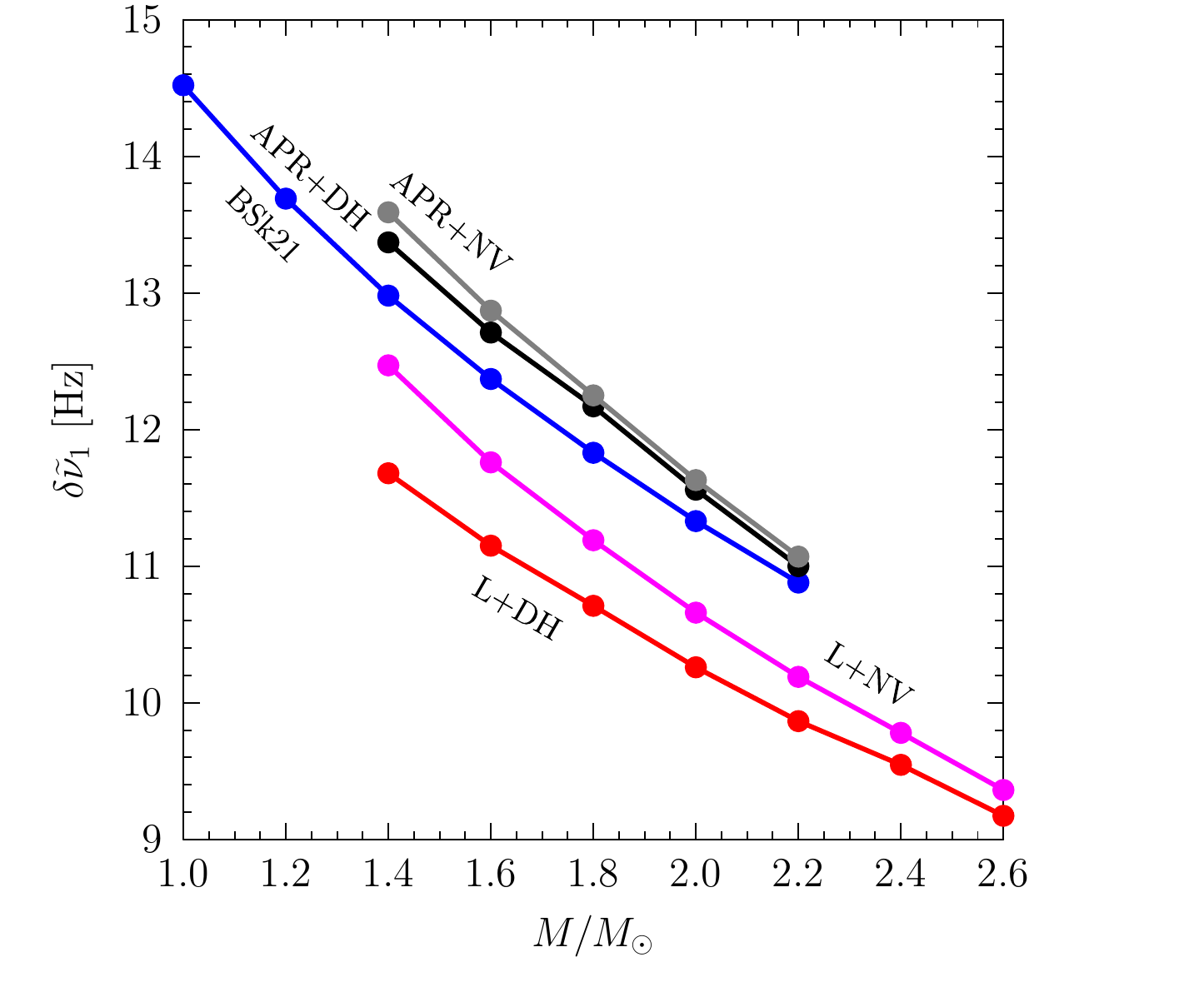}%
	\hspace{5mm}
	\caption{The auxiliary oscillation frequency $\delta \widetilde \nu_1$
		for the same conditions as in Fig.\ \ref{f:nu20(m)}.	  
	}
	\label{f:dnu1(m)}
\end{figure}

\begin{figure}
	\includegraphics[width=0.5\textwidth]{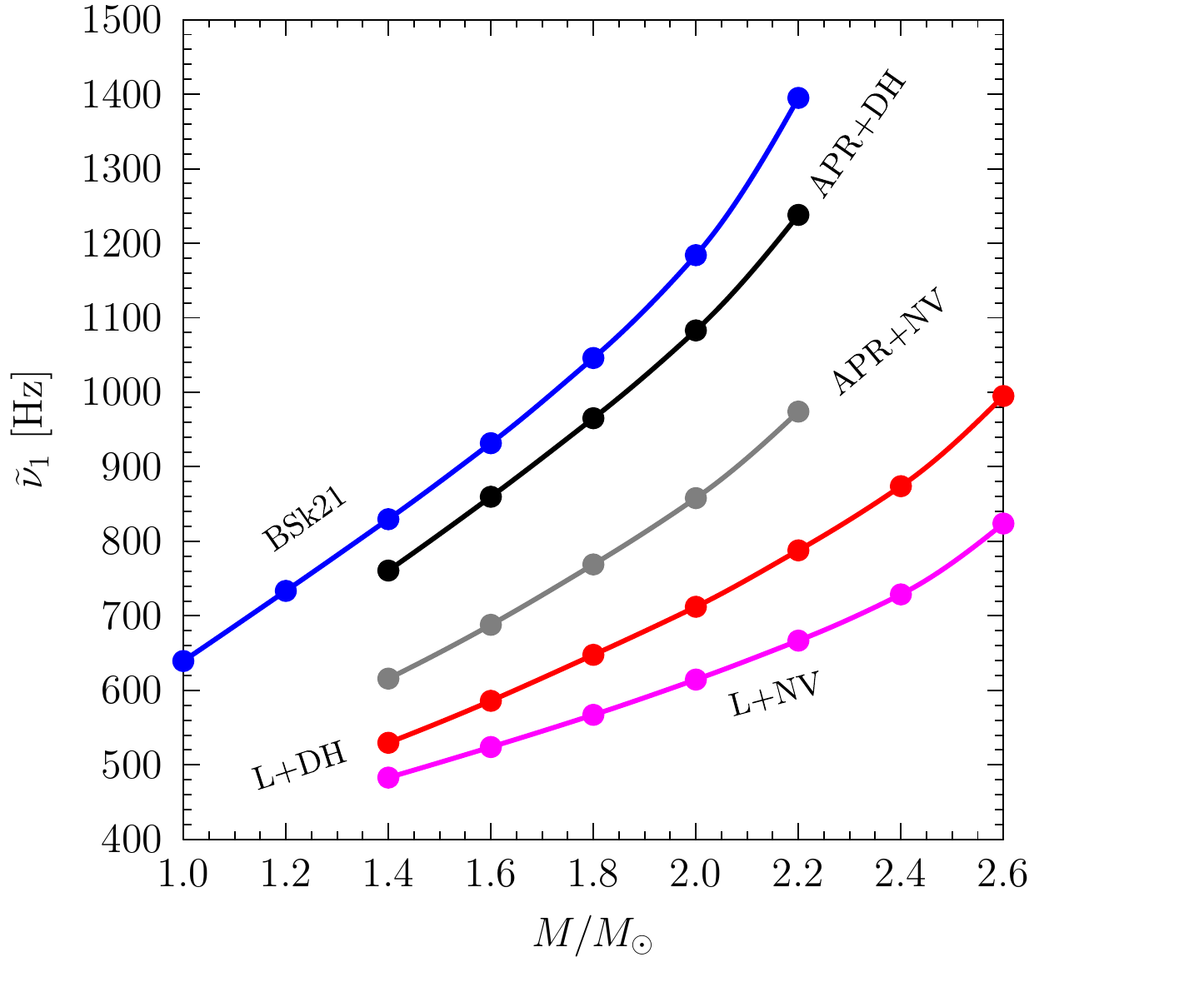}%
	\hspace{5mm}
	\caption{The auxiliary oscillation frequency $\widetilde \nu_1$
		for the same conditions as in Figs.\ \ref{f:nu20(m)} and
		\ref{f:dnu1(m)}.
	}
	\label{f:nu_1(m)}
\end{figure}

Table \ref{tab:sotani} lists the models used by \citet{2007Sotani}.
The first three columns give EOS (core+crust), mass and radius of these neutron star models. Column 4 gives the lowest fundamental frequency
$\nu_{20}$; it is obtained by
fitting the values of $\nu_{\ell 0}$ ($2 \leq \ell \leq 10$) from table 2 of
\citet{2007Sotani} by equation (\ref{e:nu-ell0}). The fits turn out
to be excellent for all models, confirming the validity of  {	
 (\ref{e:nu-ell0})}. 
Typical fit errors are about $\sim 0.001$.
These fits allowed us to find a typo in
table 2 for the L+DH EOS at $M=1.4\,\msun$ and $\ell=5$.
One should have $\nu_{50}$=57.0 instead of 60.0.

Columns 5 and 6 present the auxiliary frequencies $\widetilde\nu_1$
and $\delta \widetilde \nu_1$ obtained by fitting the 
frequencies $\nu_{\ell 1}$ at $2 \leq \ell \leq 10$  {for}  $n=1$
(table 3 of \citealt{2007Sotani})  { with equation (\ref{e:generalnua})). 
Again, the fits are fairly accurate, confirming the validity of
(\ref{e:generalnua}) at $n=1$}. 

Unfortunately, we cannot extract
detailed information on  $\widetilde\nu_n$
and $\delta \widetilde \nu_n$ with $n \geq 2$ from the data
of \cite{2007Sotani} (except for the approximate scaling
$\nu_n \propto  {n}$ mentioned in Section \ref{s:BSkAllMasses}). 
Nevertheless, we can  {check}
the dependence of $\nu_{20}$, $\widetilde \nu_1$ and
$\delta \widetilde \nu_1$ (collected in Table \ref{tab:sotani})
on neutron star mass using the scaling equations (\ref{e:Fitdnuln})
and (\ref{e:Fitnuln}) (as in Section \ref{s:bsk} for the BSk21 EOS).
Again, taking models with different masses but one EOS,
the scaling relations turn out to be very accurate. For the four
EOSs (APR+NV, APR+DH, L+NV, and L+DH) the 
fit parameters are presented in Table \ref{tab:sotani1}
(supplemented by the data for the BSk21 EOS from Table \ref{tab:main2}).
The dependence of $\nu_{20}$, $\delta \widetilde\nu_1$
and $\widetilde \nu_1$ on stellar mass for these five EOSs
is plotted, respectively, in Figs \ref{f:nu20(m)}, \ref{f:dnu1(m)}
and \ref{f:nu_1(m)}. The curves are seen to be similar to
those for the BSk21 EOS, confirming the  {correctness} of the
general description of torsional vibration modes 
(Section \ref{s:freq}). While fitting the data presented
in Table \ref{tab:sotani1}, a misprint in table 1 of
\cite{2007Sotani} has been found: the radius of the
$2 \msun$ neutron star with the L+DH EOS should
be equal to $R=14.82$ km.

\subsection{Crust bottom as a special place}
\label{s:bottom}

 {Let us stress that torsional oscillations 
of non-magnetic spherical neutron stars, discussed
above, have been computed 
neglecting some effects. 
These effects occur mainly
near the bottom of the inner crust; it is
as a special place to control torsional oscillations.}

First of all, nuclear physics is uncertain there, mostly due to uncertain density dependence of the
nuclear symmetry energy. It affects the fraction
of protons in the matter and the shear modulus $\mu$.
 {Secondly, $\mu$ is also affected by the effects of finite temperature (which
were first considered by \citealt{1991Strohmayer}). Thirdly,} one should mention superfluidity of quasi-free neutrons 
in the inner crust, which reduces the enthalpy density 
$P+\rho c^2$ of the matter ( {the} inertial mass density 
involved in torsional vibrations). Such effects were studied,
for instance,
by \citet{2012Sotani,2013Sotani,2013aSotani,2015Sotani,2016Sotani,2016aSotani}.
  {Note also that 
$\mu$ can be influenced by 
plasma screening of Coulomb interaction 
between atomic nuclei (e.g. \citealt{1991Strohmayer},
\citealt{2015Baiko}, \citealt{2022Chugunov})
and by finite sizes of the nuclei (as analyzed by \citealt{2022Sotani} on the
level of intuitive estimates and by \citealt{2022Zemlyakov} on reliable theoretical basis). The respective changes of oscillation frequencies appear noticeable and contain information on nuclear symmetry energy, superfluidity, neutron star mass, radius and compactness.}

 {All these effects can 
easily be included in the theoretical description of 
Sections \ref{s:freq}, \ref{s:bsk} and \ref{s:EOS} in a
unified manner. Appropriate fitting of oscillation frequencies
could be performed;  {it can be simpler and more natural} than the fitting proposed
in the cited publications (e.g., in \citealt{2016Sotani}).}

Additional complications can be introduced by a layer of
nuclear pasta (e.g. \citealt{HPY2007} and references therein) 
which contains exotic,
highly non-spherical nuclear clusters. It can appear   {or not
appear, depending on assumed nuclear interaction model in the layer
between the bottom of the crust of spherical nuclei and
the neutron star core}. \citet{2017aSotani,2017Sotani,2018Sotani,2019Sotani} performed
calculations of torsional oscillations in
the presence of the nuclear pasta layers. They used the standard
model of nuclear pasta as a sequence of spherical shells containing
phases of cylindrical nuclear structures, nuclear plates,
inverted cylinders, and inverted spheres. The crucial question 
is if such structures can be described in the approximation of
isotropic solid, with some effective shear modulus $\mu(r)$?
According to \citet{2019Sotani}, the dependence $\mu(r)$ within the nuclear pasta contains strong jumps. They complicate
calculating torsional oscillations. 

Another problem
is that, aside of the standard model of nuclear pasta, there are
many other  {models} which do not confirm stratification of
different pasta layers but predict mixtures of various phases
(e.g., glassy quantum nuclear plasma of \citealt{2022Newton}).
In the absence of well established model of nuclear pasta, its
effect on torsional oscillations remains uncertain. 

It seems that the uncertainties of many parameters at the
crust bottom can bias theoretical interpretation
of pure torsional oscillations of non-magnetic stars. They will equally bias magneto-elastic oscillations
of magnetars.

\section{Bursts, QPOs and  {their} interpretation}
\label{s:bursts+QPOs}

It is widely thought that oscillations of neutron stars can be triggered
by powerful bursts and superbursts occurring within the stars. The best candidates would be superbursts  {of} neutron stars
which enter compact X-ray binaries with low-mass companions.
It is known that these neutron stars do not possess strong
magnetic fields. Superbursts might generate standard torsional oscillations of non-magnetic crust.  

Superbursts  
(e.g., \citealt{2017Zand,2017Galloway})
are rare events. 
They are thought to be powered by explosive burning of carbon that is produced during nuclear evolution of accreted matter. Carbon can survive in the crust to densities $\rho \sim 
(10^7-10^{10})$ \gcc\ and then  {explode} (e.g.
\citealt{2012Altamirano,2012Keek}).  {Unfortunately},
no oscillations related to superbursts have been
observed so far.

Currently, the main attention is paid to QPOs observed in the afterglow
of flares of SGR 1806--20, SGR 1900+14, and SGR J1550--5418
(see, e.g., \citealt{2005Israel,2006Watts,2011Hambaryan,2014Huppen,
2014Huppenkothen,2015Mereghetti,2017KasB,2018Pumpe}). 
The frequencies of these QPOs  {fall} in the same range (from
$\sim 10$ Hz to a few kHz), as the expected frequencies
of torsional oscillations of non-magnetic stars. The magnetic
fields of SGRs, $B \sim 10^{15}$ G, are sufficiently
strong to affect oscillations of these stars. 

These QPOs are interpreted in two ways. The first way is to use the theory
of torsional oscillations in non-magnetic spherical stars
(e.g. \citealt{2007Sotani,2012Sotani,2013aSotani,2013Sotani,2016Sotani,2017aSotani,
2017Sotani,2018Sotani,2019Sotani}). In particular, these 
interpretations include tuning of frequencies with  the effects at the crust bottom (Section \ref{s:bottom}).

Other QPO interpretations are based on the theory 
of oscillations of magnetized  neutron stars 
(e.g., \citealt{2006Levin,2007Levin,2006Glampedakis,2007Sotani,2009CD,2009Colaiuda,2011Colaiuda,2012Colaiuda,2011vanHoven,2012vanHoven,2011Gabler,
2012Gabler,2013Gabler,2013Gabler1,2016Gabler,2018Gabler,
2014Passamon,2016Link}). A summary of this theory
can be found in \citet{2016Gabler,2018Gabler}. The theory 
predicts the existence of magneto-elastic oscillations of
the crust. They  {are} torsional oscillations modified 
by magnetic field. However, in contrast to pure torsional
oscillations, the magneto-elastic oscillations are coupled to
the core and magnetosphere via Alfv\'en waves. 

In addition, the
theory predicts global oscillations of a magnetic star based
on Alfv\'en waves. In particular, they  {contain} the so called lower, upper and edge QPOs. They are mainly localized under the crust and are 
 {determined by strength and geometry of} open and closed magnetic field lines. These oscillations strongly
depend on the physics of stellar core (on EOS, superfluidity/superconductivity, magnetic field strength and
geometry). Since these properties are rather uncertain, the
predicted QPOs can be different, which complicates unambiguous
theoretical interpretation of observations. If the magnetic field at the
crust bottom is higher than a few times of $10^{15}$ G,
the elastic crust may become  {of little importance}.

It seems that all present interpretations of QPOs face the
problem of  {too} wide space of many parameters which are currently
rather uncertain. Hopefully, the solutions will converge. 

\section{Conclusions}    
\label{s:conclude}

 {Standard exact calculation of torsional oscillation frequencies  $\nu_{\ell n}$
for a spherical non-magnetic neutron star, based on equation (\ref{e:eqforY}),
is outlined in Section \ref{s:exact}. It is valid for a star with crystalline crust that is 
treated as isotropic solid described by some shear modulus
$\mu(r)$. An equivalent formulation of
the same problem is suggested in Section \ref{s:exact1}.} 

 {Based on the latter formulation,
an approximate method for finding  $\nu_{\ell n}$
is developed (Section \ref{s:freq}). An oscillation frequency $\nu_{\ell n}$ of any mode
with fixed number $n=$0, 1, 2, \ldots of radial modes but 
different orbital numbers $\ell=$2, 3, \ldots can
be presented in the form  {(\ref{e:generalnua})}, being
determined by two auxiliary frequencies, $\widetilde \nu_n$
and $\delta \widetilde \nu_n$, independent of $\ell$. The approximate equation (\ref{e:generalnua})
appears virtually exact, at least for not too large
$\ell$. It predicts a very simple $\ell$-dependence of
$\nu_{\ell n}$ that gives a selfsimilarity relation for a star of fixed mass $M$ and EOS.}

 {For fundamental oscillations ($n=0$, {Section \ref{s:n=0}}), {$\widetilde \nu_{0}=0$} and
the spectrum (\ref{e:nu-ell0}) is determined only by {small} $\delta
\widetilde \nu_0$ {(due to inefficient meridional shear-wave energy-momentum transfer)}. For ordinary torsional modes ($n>0$, {Section \ref{s:ordinary}}),
both auxiliary frequencies contribute to $\nu_{\ell n}$, with
$\delta \widetilde \nu_n \ll \widetilde \nu_n$ (and $\delta
\widetilde \nu_n \sim \delta \widetilde \nu_0$). 
Higher auxiliary frequencies {$\widetilde\nu_n$} are associated with more intensive
radial
shear wave motions.}

 {Section \ref{s:bsk} is devoted to neutron stars of different $M$ with the BSk21 EOS, as an example. In particular, simple selfsimililar fit equations 
(\ref{e:Fitdnuln}) and (\ref{e:Fitnuln}) for
auxiliary frequencies $\delta \widetilde  \nu_{n}$ and  $\widetilde \nu_{n}$ are proposed, as functions of $M$. They appear to be very accurate
for the BSk21 EOS. Section \ref{s:previous} demonstrates that they are equally accurate 
for other EOSs considered previously by \citet{2007Sotani}.} 
  
 {One can generate a set of values $\widetilde \nu_n$
and $\delta \widetilde \nu_n$ (like in any line of Table \ref{tab:main1}) 
for a neutron star of fixed mass $M$ and calculate 
oscillation frequencies {$\nu_{\ell n}$ using equation (\ref{e:generalnua}).
Moreover, one can take tables of $\widetilde \nu_n$
and $\delta \widetilde \nu_n$ (e.g. from Table \ref{tab:main1})
for a range of $M$ (at a fixed EOS)
and interpolate the these values as functions of $M$ using equations (\ref{e:Fitnuln})
and (\ref{e:Fitdnuln}).} In this way one gets a simple and compact
description ({like} Table \ref{tab:main2}) of torsional oscillation
frequencies for all stars with given EOS and microphysics of dense
matter. The same can be done for calculating vibration energies.}

 {Section \ref{s:flatspace crust} shows that torsional
oscillation frequencies can be accurately calculated 
using the flat space-time approximation. It is expected that
this approximation can be accurate in studying magneto-elastic
oscillations in the crust of magnetized neutron stars. It
would greatly simplify semi-analytic consideration
of such oscillations, at least at not too
high magnetic fields. This would be useful for a firm search
of magneto-elastic oscillations in the spectra of magnetar's QPOs.}    

 {Section \ref{s:bottom} outlines current uncertainties
of microphysical functions ($\rho(r)$, $P(r)$ and $\mu(r)$), 
which
enter the basic equation (\ref{e:eqforY}), at the crust bottom.
These uncertainties seem significant; much work is required to reduce them.}

 {Finally,  Section
\ref{s:bursts+QPOs} discusses prospects of applying the obtained results
for 
interpretation of observations.}

\section*{Acknowledgments}
 {I am very grateful to the anonymous referee for constructive critical comments.}
The work ( {except for Section \ref{s:flatspace crust}}) was supported by the Russian
Science Foundation (grant 19-12-00133 P). 

\section*{Data availability}
The data underlying this article will be shared on
reasonable request to the authors.

\bibliographystyle{mnras}


\end{document}